\documentclass[ reprint,groupedaddress, amsmath,amssymb, aps, longbibliography]{revtex4-1}

\usepackage[colorlinks=true,linkcolor=blue,urlcolor=blue,citecolor=blue]{hyperref}
\usepackage{overpic}
\usepackage{upgreek}
\usepackage{physics}
\usepackage{graphicx}
\usepackage{dcolumn}
\usepackage{bm}
\usepackage{outlines}
\usepackage[dvipsnames]{xcolor}
\usepackage{gensymb}
\usepackage{subcaption}
\usepackage{caption}
\usepackage{tikz}  
\usepackage{comment}
\usepackage{booktabs} 

\usepackage{pdfpages}
\makeatletter
\AtBeginDocument{\let\LS@rot\@undefined}
\makeatother

\newcommand{\mitll}{MIT Lincoln Laboratory, Lexington, MA 02421, USA}

\begin{document}

\title{Mitigation of Magnetic Flux Trapping in Superconducting Electronics Using Moats}
\author{Rohan T.~Kapur}
\email{rohan.kapur@ll.mit.edu}
\author{Sergey K.~Tolpygo}
\email{sergey.tolpygo@ll.mit.edu}
\author{Alex Wynn}
\author{Pauli Kehayias}
\author{Adam A.~Libson}
\author{Collin N.~Muniz}
\author{Michael J.~Gold}
\author{Justin L.~Mallek}
\author{Danielle A.~Braje}
\author{Jennifer M.~Schloss}
\affiliation{\mitll}

\date{\today}

\begin{abstract}
Magnetic flux (vortex) trapping remains a major obstacle to very large scale integration in superconducting electronics. Moats --- etched regions in circuit layers placed in ground planes and around critical circuitry --- offer a simple passive approach to sequester flux. Here, we systematically examine the effectiveness of moat arrays in superconducting niobium films as a function of geometry (size, shape, and density) and background magnetic field. By measuring the vortex expulsion field, we estimate the flux saturation number and flux trapping temperature for a range of geometries. We find that many moat designs effectively sequester flux in magnetically shielded environments ($< 1$ $\upmu$T), with high-aspect-ratio rectangular ``slit" moats providing the strongest mitigation at minimal area cost. However, our measurements show that moats alone do not eliminate flux trapping in non-ideal films, as vortices can preferentially pin at material defects. These results provide design guidance for flux mitigation in superconducting integrated circuits and highlight the need for combined optimization of circuit geometries and materials.
\end{abstract}
\maketitle
\section{Introduction}
Superconductor classical electronics remain among the most compelling alternatives to traditional complementary metal-oxide-semiconductor (CMOS) electronics, offering orders of magnitude improvements in energy efficiency and clock speeds~\cite{SCEreviewVanDuzer, SCEreviewSpringer, SCEreviewIEEE2024}. Despite these advantages, the commercial viability of superconductor electronics (SCE) remains limited by scalability challenges, with current integration densities being 3-4 orders of magnitude lower than those achieved in CMOS technologies~\cite{shregOld, nagasawa_sce_scaling, ac_power_sfq, Herr_2015, Mutsuo_HIDAKA20212020SUI0002, ayala_monolithic_infra}. One significant factor constraining scaling is magnetic flux trapping, a phenomenon that is unique to SCE and distinct from conventional scaling barriers such as heat dissipation and fabrication feature size~\cite{shregNew, washington1982observation, semenov_flux_trapping_2009}.

SCE typically use thin-film Nb, NbN, and NbTiN, all type-II superconductors, for their relatively high superconducting critical temperature, $T_c$, large critical current densities and high critical magnetic fields, good chemical and mechanical stability, and processing advantages relative to other superconductors~\cite{likharev2012superconductor, tolpygo2016superconductor, tahara2002superconducting, tolpygo2023progress, nieto2023flexible,pokhrel2024nbtin}. When a type-II superconducting film is subjected to a residual magnetic field, $B_r$, larger than its critical magnetic field, $B_{c1}$, or is cooled through the superconducting critical temperature in the presence of any residual magnetic field ($B_r\neq 0$), it can trap quantized magnetic flux in the form of vortices, which can interfere with and even completely compromise circuit operation. Various strategies have been proposed and implemented to mitigate flux trapping, including operating in a magnetically shielded environment, applying thermal gradients to drive flux away from critical circuitry~\cite{veshchunov2016optical, geng1992sweeping}, defluxing and vortex ratcheting with AC fields~\cite{ratchetEffect, ac_deflux_squid}, and using etched structures in superconducting films to sequester flux away from circuitry. The latter includes slots, slits, and circular or square holes surrounding or dispersed in the circuitry and ground planes, which are often referred to as moats or antidots~\cite{semenov2016moats,IBMmoats,ssmMoats1995,fourie2021experimental, colauto2020controlling}. However, a comprehensive, universally effective solution for flux trapping in circuits has yet to be found. 

Introducing moats into superconducting circuits, if fully effective, would constitute the simplest practical solution to flux trapping, requiring no active circuitry to function. Whereas many prior studies of moat arrays have focused on increasing vortex pinning and high-field behavior~\cite{berdiyorov2006novel, vestgaarden2012mechanism, priour2004vortex, raedts2004flux, berdiyorov2006superconducting, PhysRevB.74.174512, silhanek2005enhanced}, we evaluate the effectiveness of moat arrays in maintaining a vortex-free (Meissner) state of films cooled in low background magnetic fields ($B_r \leq$ 10 $\upmu$T), typical for superconductor electronics operation. The goal is to engineer moat arrays that achieve strong flux mitigation, while also minimizing the fractional area occupied by moats. 

We analyze the effectiveness of moats at mitigating trapped flux by measuring the vortex expulsion field as a function of moat number density and geometry in single-layer Nb films typical of SCE ground planes. Ground planes are generally the dominant flux-trapping regions in SCE circuits due to their large, continuous area~\cite {shregNew, shregOld, ac_power_sfq, ssmMoats1995}. The vortex expulsion field, $B_{\mathrm{exp}}$, is the maximum background magnetic field in which an ideal, defect-free superconducting structure can be cooled through $T_c$ without forming vortices, and thus represents an approximate upper bound on $B_r$ before vortex-induced effects on circuit operation must be considered. Ideally, all circuit elements would be designed with expulsion fields larger than the operating background field, rendering flux trapping irrelevant.

We show that many existing moat designs trap the majority of vortices in $B_r\lesssim$ 10 $\upmu$T with rectangular ``slit" moats (aspect ratio $\geq30$) offering the most effective geometry for strong flux mitigation at a small area footprint. Our results indicate, however, that the effectiveness of moats in non-ideal films is fundamentally constrained by vortex pinning, wherein magnetic flux becomes trapped on microscopic defects within the superconducting film unable to reach moats. This vortex trapping on defects can occur even at very low background fields ($B_r\lesssim$ 1 $\upmu$T), limiting the ability of moats to fully eliminate trapped flux and indicating the need for additional mitigation strategies.
\section{Results}
\label{sec:results}
\subsection{Square Moat Arrays}
\begin{figure}[htbp]
  \centering
  \captionsetup{font=small}

  \begin{overpic}[width=\columnwidth, trim=35pt 0pt 0pt 0pt, clip]{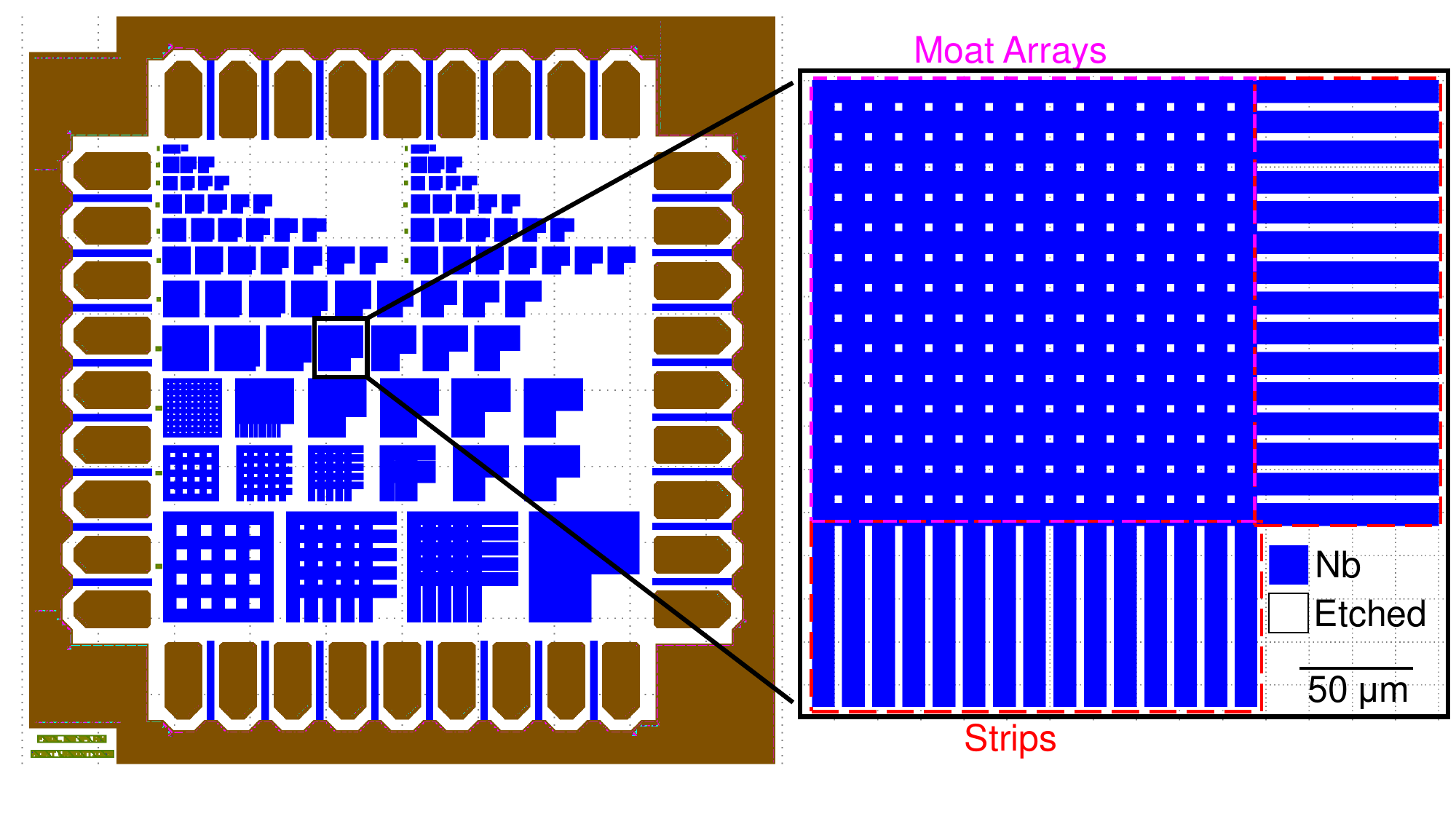}
    \put(-8,52){\footnotesize\textbf{(a)}}
  \end{overpic}

  \vspace{0.5em}

  \begin{overpic}[width=1.02\columnwidth, trim=0pt 0pt 0pt 0pt, clip]{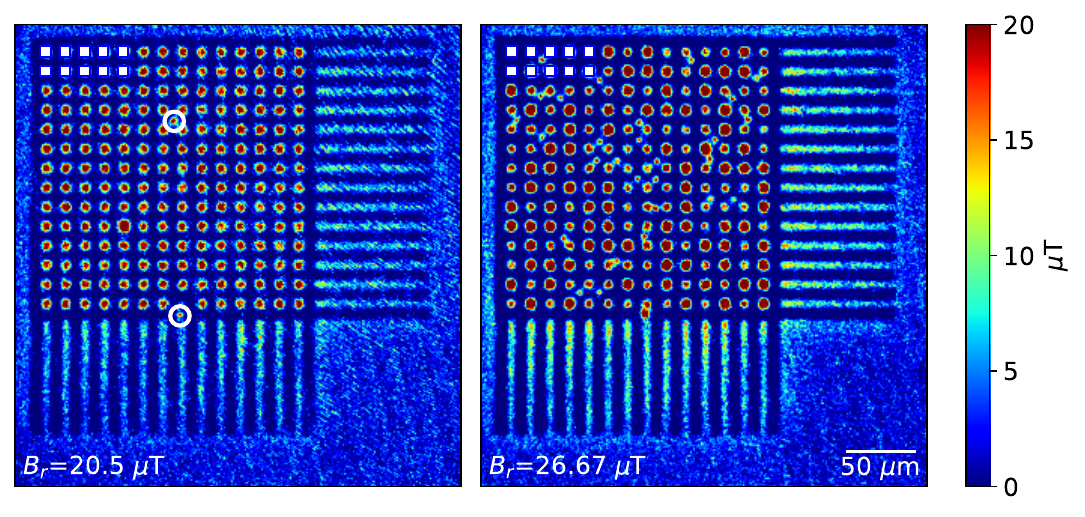}
    \put(-8,42){\footnotesize\textbf{(b)}}
  \end{overpic}

  \caption{
  (a) Layout of test chip used to characterize flux trapping in square moat (antidot) arrays. The spacing between moats varies across rows (0.5--80~$\upmu$m), while the moat side length varies across columns (1--80~$\upmu$m). The highlighted region shows the array with moat side length $a = 4~\upmu$m and spacing $s = 10~\upmu$m.
  (b) Magnetic images of the $a = 4~\upmu$m, $s = 10~\upmu$m array at $B_r < B_{\mathrm{exp}}^{\mathrm{meas}}$ (left) and $B_r > B_{\mathrm{exp}}^{\mathrm{meas}}$ (right). Vortices that first appear below $B_{\mathrm{exp}}^{\mathrm{meas}}$ are circled in white (left). Across temperature cycles and background fields, vortices consistently nucleate at these locations.}
  \label{fig:square_moat_array_fig}
\end{figure}

To systematically characterize moat effectiveness, we measured $B_{\mathrm{exp}}$ of Nb square moat arrays --- thin-film Nb squares patterned with two-dimensional square lattices of square holes (moats) --- as a function of moat side length $a_x = a_y = a$ and spacing $s_x = s_y = s$ (pitch $p_x = a_x + s_x$, $p_y = a_y + s_y$, and $p = a + s$) (see Fig.~\ref{fig:square_moat_array_fig}). These arrays are representative of typical ground-plane architectures in superconducting integrated circuits, which are composed of large, periodic moat arrays patterned into the electrically connected circuit ground planes~\cite{shregNew}. 

For this study, we fabricated $5~\mathrm{mm} \times 5~\mathrm{mm}$ test chips containing $d=200$ nm thick Nb films patterned with square arrays of square moats with $a$ ranging from 0.5 to 80~$\upmu$m and $s$ ranging from 1 to 80 $\upmu$m, shown in Fig.~\ref{fig:square_moat_array_fig}(a).
Nb films were deposited on 200-mm Si wafers coated with 100 nm PECVD SiO$_2$, using Nb deposition and etching parameters identical to the MIT LL SFQ5ee process \cite{SFQ5ee}. 

We characterized $B_{\mathrm{exp}}$ of moat arrays with $a$ and $s$ ranging from 0.5 to 30 $\upmu$m and 4 to 40 $\upmu$m, respectively, and $a\leq s$. The measurements were performed by field cooling the arrays in various $B_r$ and taking magnetic images of the resultant flux distribution. Magnetic images were acquired using a cryogenic widefield nitrogen-vacancy (NV) diamond microscope, which provides quantitative maps of the out-of-plane magnetic field with micron-scale spatial resolution and sensitivity sufficient to resolve individual vortices over millimeter-scale areas from room temperature down to $\sim 5$ K~\cite{qswift_apparatus_paper}. From the images, the vortex areal density, $n_v$, was extracted as a function of $B_r$. 

\begin{figure*}[htbp]
  \centering
  \captionsetup{font=small}
  
  \begin{overpic}[width=0.48\textwidth]{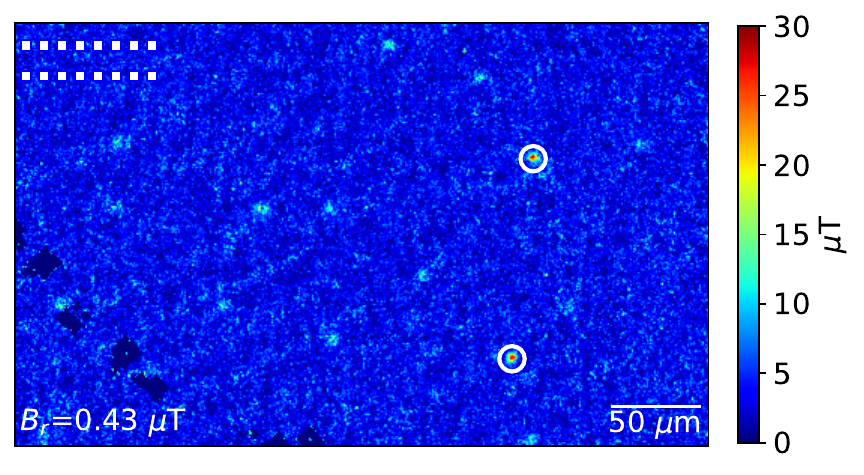}
    \put(-6,52){\footnotesize\textbf{(a)}}
  \end{overpic}
  \hspace{0.02\textwidth}
  \begin{overpic}[width=0.48\textwidth]{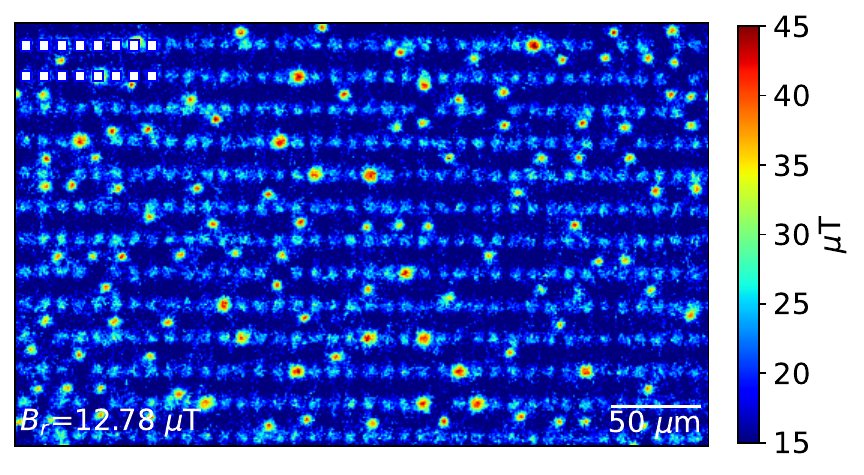}
    \put(-6,52){\footnotesize\textbf{(b)}}
  \end{overpic}

  \vspace{0.75em}

  \begin{overpic}[width=0.48\textwidth]{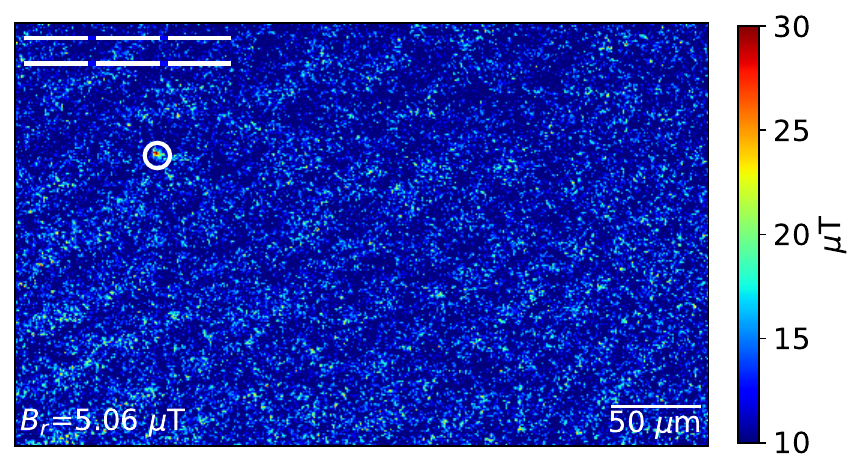}
    \put(-6,52){\footnotesize\textbf{(c)}}
  \end{overpic}
  \hspace{0.02\textwidth}
  \begin{overpic}[width=0.48\textwidth]{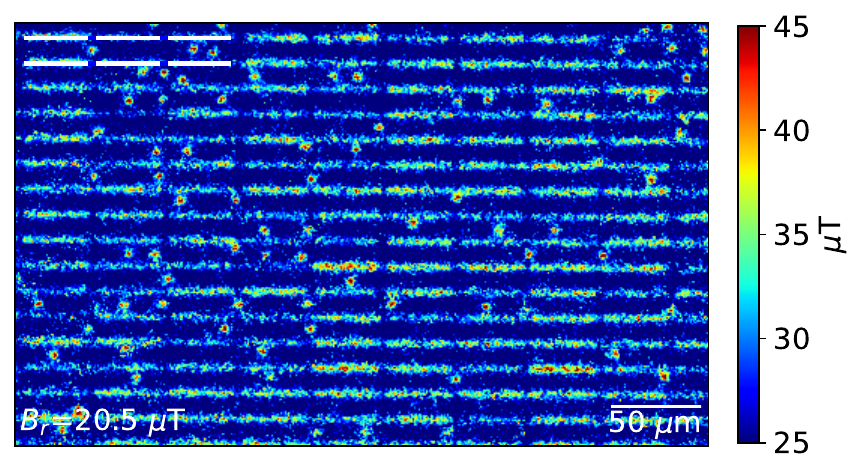}
    \put(-6,52){\footnotesize\textbf{(d)}}
  \end{overpic}

  \caption{
  (a)--(b) Example images of a moat array with $a_x=a_y=4~\upmu$m, $s_x=6~\upmu$m, and $s_y=14~\upmu$m. (c)--(d) Example images of a moat array with $a_x=36~\upmu$m, $a_y=1~\upmu$m, $s_x=4~\upmu$m, and $s_y=13~\upmu$m. In each image, the first few moats in the upper left are marked (white squares in (a)--(b), white stripes in (c)--(d)), and the first vortices to appear in the film are circled in white. These vortices reproducibly emerge at the same locations across temperature cycles, indicating nucleation and/or pinning at defect sites.
}
  \label{fig:asym_examples}
\end{figure*}

\begin{figure}[htbp]
  \centering
  \captionsetup{font=small}

  \begin{overpic}[width=\columnwidth,trim=0pt 0pt 0pt 0pt,clip]{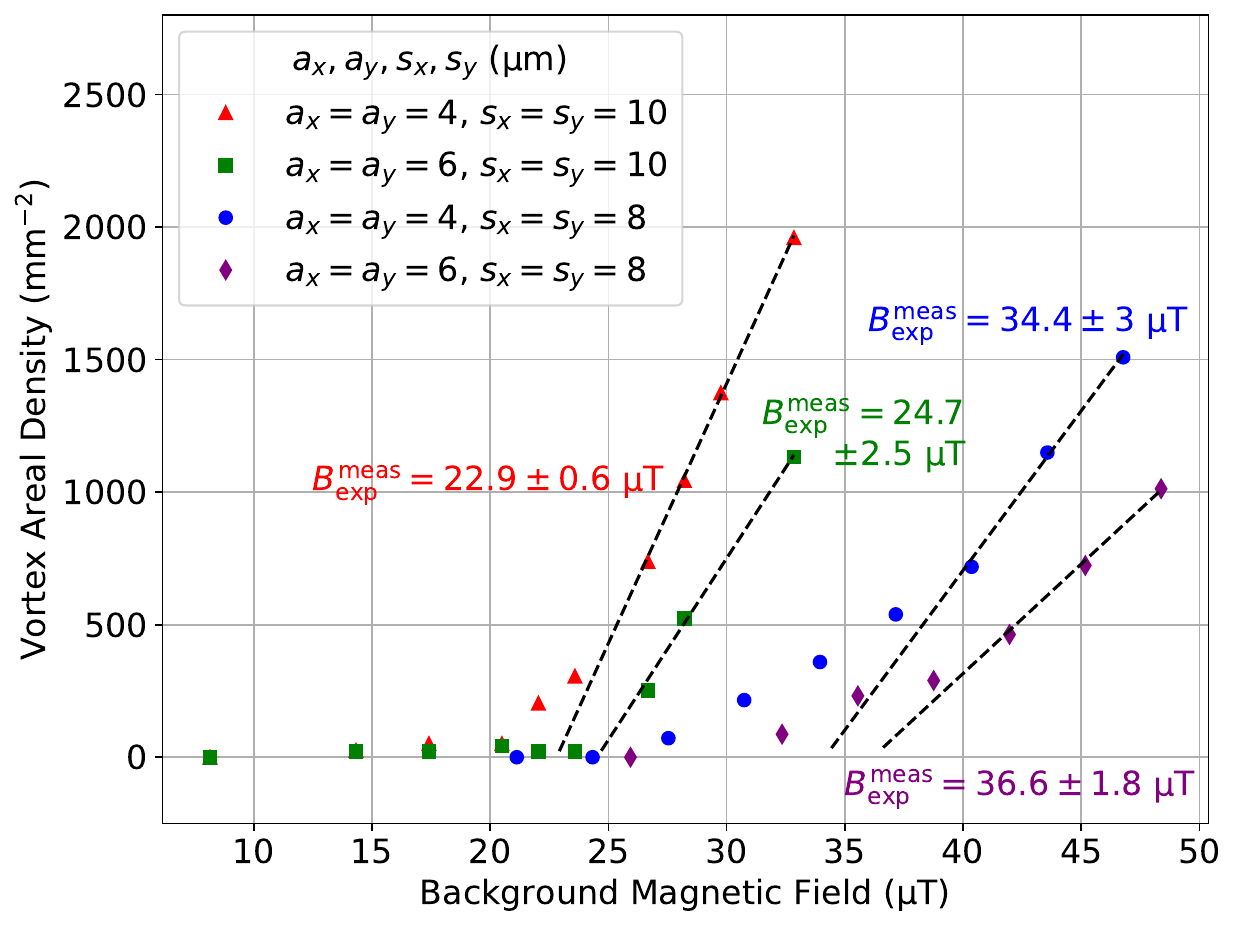}
    \put(-2,75){\footnotesize\textbf{(a)}}
  \end{overpic}%
  \hfill

  \begin{overpic}[width=\columnwidth,trim=0pt 0pt 0pt 0pt,clip]{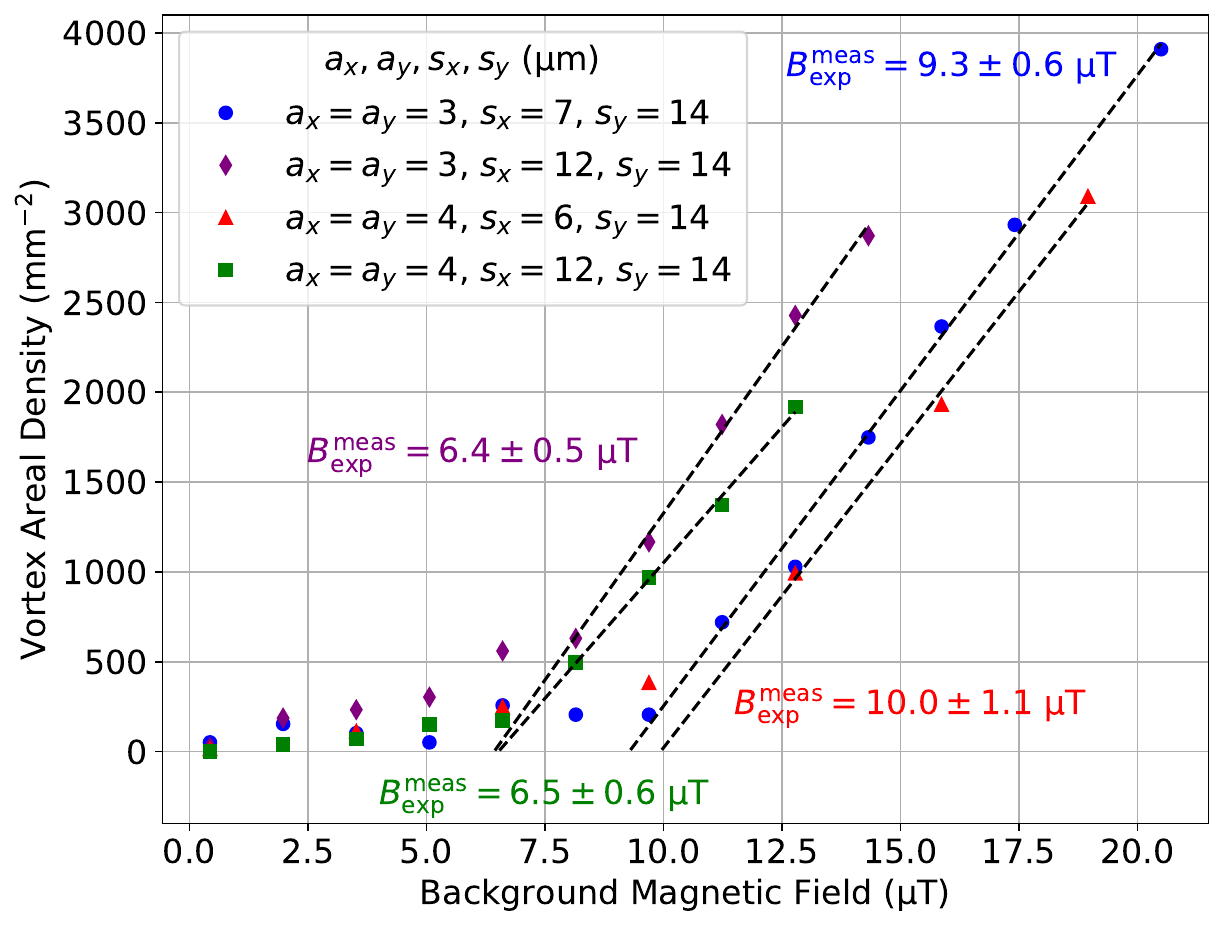}
    \put(-2,75){\footnotesize\textbf{(b)}}
  \end{overpic}%
  \hfill

  \begin{overpic}[width=\columnwidth,trim=0pt 0pt 0pt 0pt,clip]{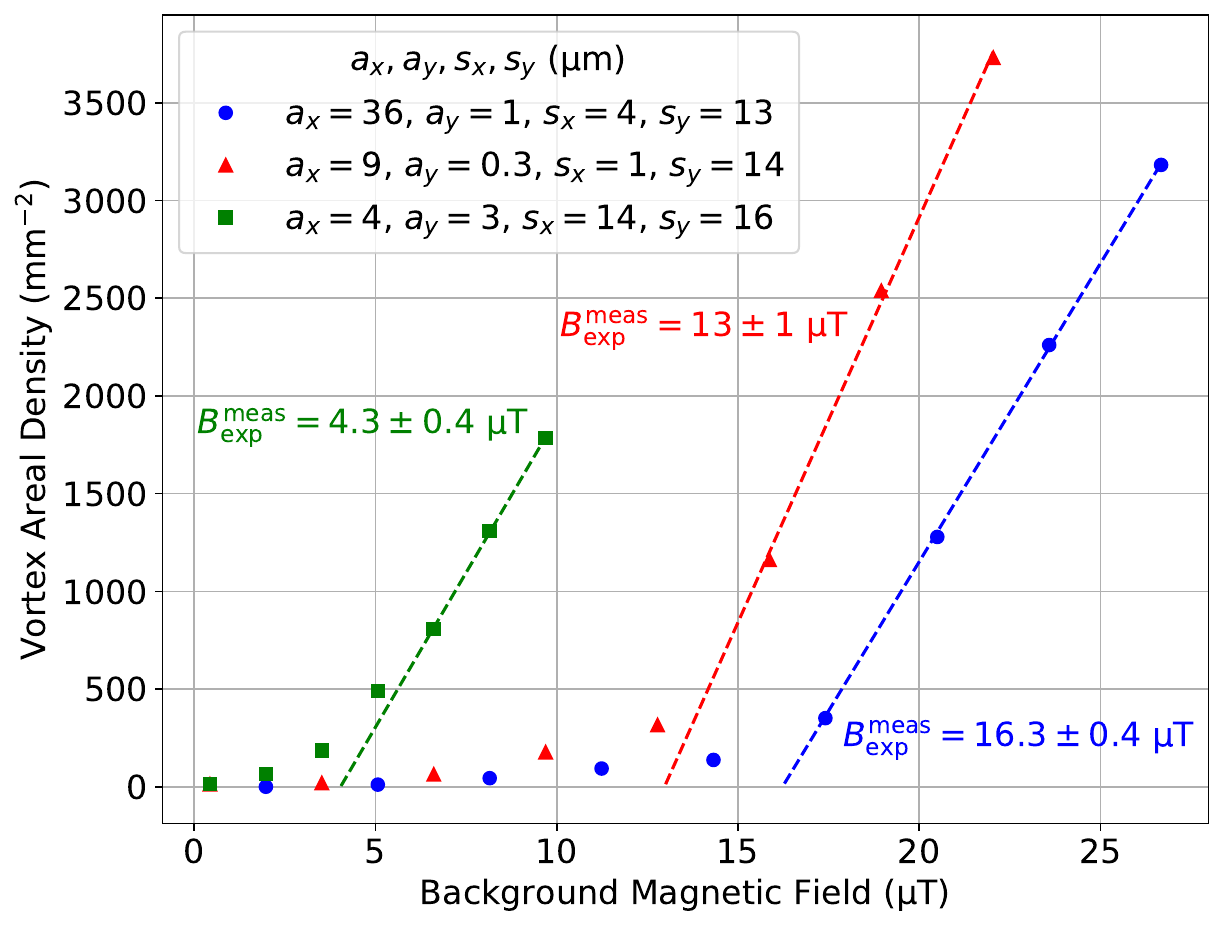}
    \put(0,75){\footnotesize\textbf{(c)}}
  \end{overpic}%

  \caption{
  Example expulsion-field measurements for (a) square moat arrays (\(a_x=a_y\), \(s_x=s_y\)), (b) square arrays with anisotropic spacing (\(a_x=a_y\), \(s_x\neq s_y\)), and (c) rectangular arrays with anisotropic spacing (\(a_x\neq a_y\), \(s_x\neq s_y\)). For each array, the vortex areal density \(n_v(B_r)\) is obtained by imaging the vortex distribution at each \(B_r\), and \(B_{\mathrm{exp}}^{\mathrm{meas}}\) is extracted from the onset of a linear regime \(n_v(B_r)=m\Phi_0^{-1}(B_r-B_{\mathrm{exp}}^{\mathrm{meas}})\) by extrapolating to \(n_v=0\). Slope \(m\) depends on the moat array geometry and is discussed in Appendix~\ref{subsec:exp_field_slope}. We find \(B_{\mathrm{exp}}^{\mathrm{meas}}\) depends most strongly on moat spacing, with a smaller but nontrivial dependence on moat size and shape. In all cases, \(n_v \neq 0\) at \(B_{\mathrm{exp}}^{\mathrm{meas}}\), as vortices appearing below this field nucleate and/or pin reproducibly at defect sites in the film.
  }

  \label{fig:exp_field_measurements}
\end{figure}

From the analysis of magnetic field images, examples of which are shown in Fig.~\ref{fig:square_moat_array_fig} and~\ref{fig:asym_examples}, we observed two characteristic fields, $B_1$ and $B_{\mathrm{exp}}^{\mathrm{meas}}$. The first field, $B_1$, is defined as the background field $B_r$ above which vortices first appear in the film outside the moats. Across multiple cooldowns, these first few vortices always appear at the same locations in the film, indicating that they either nucleate or become pinned at these locations. These sites likely correspond to defects in the film acting as strong vortex pinning centers. \textcolor{black}{The areal density of these pinned vortices slowly increases with increasing $B_r$, displaying a convex $n_v(B_r)$ function which eventually changes to a linear dependence as shown in Fig.~\ref{fig:exp_field_measurements}.}

We define the measured expulsion field, $B_{\mathrm{exp}}^{\mathrm{meas}}$, of each moat array as the onset of a linear increase in vortex areal density with $B_r$, $n_v = m \Phi_0^{-1}(B_r - B_{\mathrm{exp}}^{\mathrm{meas}})$, where $\Phi_0 = h/2e \approx 2.07 \times 10^{-15}\,\mathrm{T \cdot m^2}$ is the magnetic flux quantum. \textcolor{black}{We observe that the slope $m$ varies between arrays from $m\approx 0.1$ to $0.8$; see Fig.~\ref{fig:high_field_moat5} in Appendix~\ref{subsec:exp_field_slope}. This indicates that, while a significant amount of flux above $B_{\mathrm{exp}}^{\mathrm{meas}}$ is trapped as vortices in the film, the moats continue to trap a fraction of the additional applied flux. Arrays with higher $B_{\mathrm{exp}}^{\mathrm{meas}}$ generally exhibit smaller slopes than those with lower $B_{\mathrm{exp}}^{\mathrm{meas}}$. This indicates that more efficient arrays with higher $B_{\mathrm{exp}}^{\mathrm{meas}}$ continue trapping a larger fraction of the applied flux inside the moats in the $B_r > B_{\mathrm{exp}}^{\mathrm{meas}}$ regime. A more detailed discussion of the $n_v(B_r)$ dependence is provided in Appendix~\ref{subsec:exp_field_slope}.}

Hereafter, we will analyze $B_{\mathrm{exp}}^{\mathrm{meas}}$ of the measured moat arrays as it likely characterizes the expulsion field of an ideal patterned film, whereas $B_1$ is more dependent on material quality, film defects, and non-idealities rather than film geometry. Example measurements of the square moat arrays can be found in Fig.~\ref{fig:square_moat_array_fig} and \ref{fig:exp_field_measurements}(a). \textcolor{black}{The extracted $B_{\mathrm{exp}}^{\mathrm{meas}}$ for all square moat arrays is given in Appendix~\ref{subsec:sq_moat_arrays}.}

\subsection{Rectangular Moat Arrays}
Square moat arrays enable direct comparison of the effects of moat spacing and size on the vortex expulsion field, but they are not necessarily commensurate with circuit design constraints or represent optimal geometries, which may be rectangular ($a_x \neq a_y$) or have anisotropic spacing ($s_x \neq s_y$). For example, SCE devices with rectangular moat arrays in ground planes have demonstrated similarly robust low-field operation while occupying less area~\cite{shregNew}. To study rectangular moat arrays, we fabricated a second test chip, under the same processing conditions as the square-moat chip, with moat arrays identical to those in the largest operational SCE circuits fabricated in the MIT LL SFQ5ee process~\cite{SFQ5ee, shregNew}. The chip contained 10 distinct rectangular moat arrays which can be classified in two groups: square moats with anisotropic spacing ($a_x=a_y$, $s_x\neq s_y$, see Fig.~\ref{fig:asym_examples}(a)-(b)) and rectangular moats with anisotropic spacing ($a_x\neq a_y$, $s_x\neq s_y$, see Fig.~\ref{fig:asym_examples}(c)-(d)). We measured the $n_v(B_r)$ dependence and vortex expulsion field of each array with the spacing $s_x/s_y$ and moat aspect ratios $a_x/a_y$ ranging from 0.07 to 0.88 and 1 to 65, respectively. As in the square moat arrays, a small number of vortices appear at $B_r < B_{\mathrm{exp}}^{\mathrm{meas}}$ in all measured rectangular arrays, consistently at the same spatial locations and likely due to film non-idealities. Example images of flux distribution in the rectangular moat arrays and expulsion field measurements are in Fig.~\ref{fig:asym_examples} and \ref{fig:exp_field_measurements}(b)-(c), while information about each measured rectangular moat array can be found in Table~\ref{tab:rect_moats}.

\begin{table}[htbp]
\centering
\footnotesize
\begin{tabular}{|c|c|c|c|c|c|}
\hline
$a_x$ ($\upmu$m) & $a_y$ ($\upmu$m) & $s_x$ ($\upmu$m) & $s_y$ ($\upmu$m) & Moat Area & $B_{\mathrm{exp}}^{\mathrm{meas}}$ ($\upmu$T) \\
\hline
3    & 3   & 7   & 14   & 5.3\%  & 9.3(6) \\
\hline
3    & 3   & 12   & 14   & 3.5\%  & 6.4(5) \\
\hline
4    & 3   & 14   & 16   & 3.5\%  & 4.3(4) \\
\hline
4    & 4   & 6   & 14   & 8.9\%  & 10.0(11) \\
\hline
4    & 4   & 12   & 14   & 5.6\%  & 6.5(6) \\
\hline
4    & 4   & 14   & 16   & 4.4\%  &  4.3(4) \\
\hline
5    & 5   & 5   & 14   & 13.2\% & 8.8(5) \\
\hline
9    & 0.3 & 1   & 14   & 1.8\%  & 13(1) \\
\hline
19.5 & 0.3 & 1 & 14   & 1.9\%  &  15.9(4)\\
\hline
36   & 1   & 4   & 13   & 6.4\%  & 16.3(4) \\
\hline
\end{tabular}
\caption{Geometries and measured expulsion fields of all measured rectangular moat arrays. The measured geometries correspond to the moat geometries used in the largest operational SCE circuits fabricated in the MIT LL SFQ5ee process~\cite{SFQ5ee, shregNew}.}
\label{tab:rect_moats}
\end{table}

\section{Discussion}
The simplest model for the vortex expulsion field assumes that all flux threading an array unit cell with area $p_x p_y$ in the normal state, $B_r p_x p_y$, is expelled into the single moat it contains upon cooling through $T_c$. If each moat can hold no more than $N_{\Phi_0}$ flux quanta, vortices will appear in the film if $B_r / (n_{\mathrm{moat}} \Phi_0) >  N_{\Phi_0}  $, which gives the expulsion field of an array as
\begin{equation}
        B_{\mathrm{exp}} = \langle N_{\Phi_0} \rangle \Phi_0 n_\mathrm{moat},
        \label{eq:Bexp_basic}
\end{equation} 
where $\langle N_{\Phi_0} \rangle$ is the average flux per moat in the array and $n_\mathrm{moat}$ is the moat areal density. For an individual moat, $N_{\Phi_0}$ is an integer dependent on the moat and array geometry; see Appendix.~\ref{subsec:isolated_moats}. 

In the measured films, $n_\mathrm{moat}$ =  $(p_x p_y)^{-1}$ because $L, W \gg p_x, p_y$, where $L$ and $W$ are the dimensions of the array. The finite film extent and material defects can modify $B^{\mathrm{meas}}_{\mathrm{exp}}$ relative to Eq.~\ref{eq:Bexp_basic}, as flux may be expelled to the film edges or trapped on defects rather than solely accommodated by the moats. Consequently, $N_{\Phi_0}$ need not be uniform across unit cells and $\langle N_{\Phi_0} \rangle$ may deviate slightly from an integer. For the measured arrays ($W \times L \geq 100 \times 100~\upmu\mathrm{m}^2$), these effects are small compared to the array-induced behavior, consistent with the observed clustering of $\langle N_{\Phi_0} \rangle$ near integer values. The influence of defects is further accounted for by our use of $B^{\mathrm{meas}}_{\mathrm{exp}}$ rather than $B_1$ (see Sec.~\ref{sec:results}).

\begin{figure}[htbp]
    \centering
    \vspace{0.5em}
    \begin{overpic}[width=0.95\linewidth]{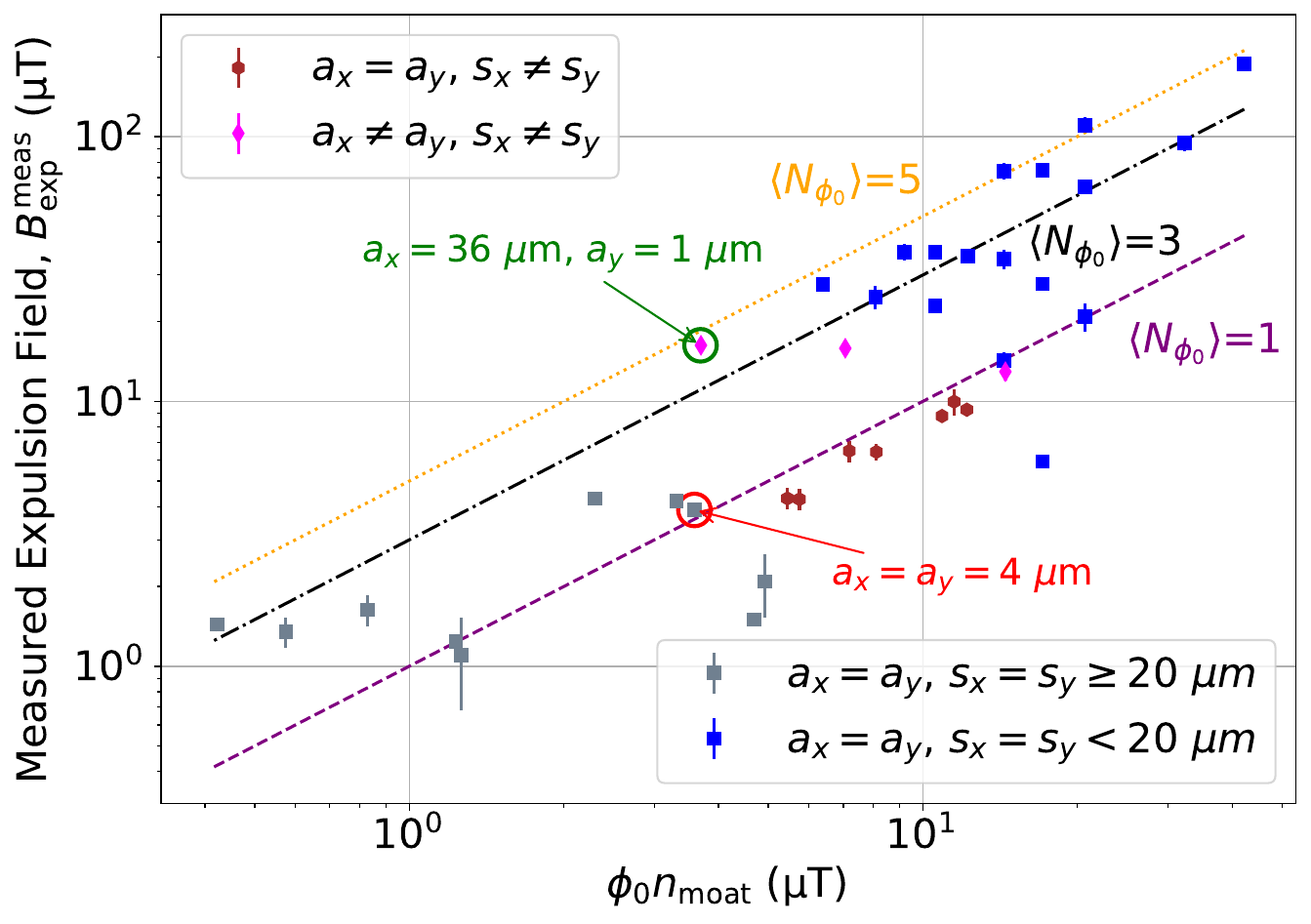}
    \end{overpic}

    \caption{Measured expulsion field vs.~$ \Phi_0n_\mathrm{moat}$. The purple, black, and orange lines indicate the expected behavior for $\langle N_{\Phi_0}\rangle =1$, 3, and 5, respectively. Square arrays with $s \geq 20~\upmu$m and arrays with anisotropic spacing agree well with $\langle N_{\Phi_0}\rangle=1$. For $s < 20~\upmu$m and highly asymmetric moats ($a_x/a_y \geq 30$), the data instead fall within $1 \leq \langle N_{\Phi_0}\rangle\leq 5$, indicating multiple flux quanta per moat before vortices enter the film. For example, at comparable moat densities, $a_x = 36$ $\upmu$m, $a_y=1$ $\upmu$m slits trap $\sim5\times$ more flux than $a = 4$ $\upmu$m squares.}

    \label{fig:Bexp_vs_basic_scaling}
\end{figure}

In Figure~\ref{fig:Bexp_vs_basic_scaling}, $B_{\mathrm{exp}}^{\mathrm{meas}}$ is plotted vs. $\Phi_0 n_\mathrm{moat}$ for the studied moat arrays. \textcolor{black}{The slope of this dependence provides an estimate of $\langle N_{\Phi_0} \rangle$ for each array, which ranges from $\langle N_{\Phi_0} \rangle = 1$ to $\langle N_{\Phi_0} \rangle = 5$.} Fig.~\ref{fig:Bexp_vs_basic_scaling} shows that, at fixed moat density, the amount of flux per moat at the expulsion field strongly depends on moat shape and size. For example, at comparable moat densities, $a_x = 36$ $\upmu$m, $a_y=1$ $\upmu$m slits trap $\sim5\times$ more flux than $a = 4$ $\upmu$m squares. The data also indicate that $\langle N_{\Phi_0} \rangle$ depends strongly on moat spacing (pitch), with dense arrays of moats trapping significantly more flux than sparse arrays. Among the measured moat arrays, slit moats appear to be the most attractive for SCE applications due to their small area footprint and ability to trap multiple flux quanta.

To better understand the expulsion field of films with periodic arrays of slits, we consider the moats in the idealized limiting case $a_x \rightarrow p_x$. In this limit, the spacing between the moats along one dimension approaches zero, $s_x \rightarrow 0$, and the initial superconducting film breaks into a set of parallel horizontal strips with width $s_y$ and spacing $a_y$. $B_{\mathrm{exp}}$ for a single long strip with width $W$ has been extensively studied both theoretically and experimentally and is given by
\begin{equation}
        B_{\mathrm{exp}} = \beta_{\mathrm{str}} \Phi_0 / W^2,
        \label{eq:Bexp_strips}
\end{equation} 
where $\beta_{\mathrm{str}}$ is a parameter that assumes different values in different theoretical models and experimentally was found to weakly depend on the superconducting film parameters~\cite{washington1982observation, likharev1971formation, bean1964surface,ybco_strip_critical_field, kuit2009vortex, initial_sc_strip_paper,qswift_apparatus_paper}.  \textcolor{black}{For instance, $\beta_{\mathrm{str}}= 3.43$ was found for Nb strips with $W\leq10$ $\upmu$m and $\beta_{\mathrm{str}}= 1.89$ for the strips with $W\geq20$ $\upmu$m~\cite{qswift_apparatus_paper}}.  If the presence of adjacent strips does not significantly affect $B_{\mathrm{exp}}$, $B_{\mathrm{exp}}$ for films with square arrays of square moats with $p-a\ll a$ should scale as 
\begin{equation}
        B_{\mathrm{exp}} = \beta \Phi_0 / s_y^2.
        \label{eq:Bexp_slits}
\end{equation} 

Two-dimensional rectangular arrays of rectangular moats can be approximated by either horizontal strips with width $s_y$ in the limit $a_x/p_x \rightarrow 1$, or vertical strips with width $s_x$ in the limit $a_y/p_y \rightarrow 1$. In the latter case, $s_x$ would replace $s_y$ in Eq.~\ref{eq:Bexp_slits}. The simplest interpolation between these two dependencies is
\begin{equation}
        B_{\mathrm{exp}} = \frac{\beta_{moat} \Phi_0}{ s_x^2 + s_y^2},
        \label{eq:Bexp_slits_2d}
\end{equation} 
where $s_x^2 + s_y^2 \equiv 2 s_{\mathrm{eff}}^2$ represents the square of the effective width - the widest part of the film between moats. 

\begin{figure}[htbp]
    \centering
    \vspace{0.5em}
    \begin{overpic}[width=0.95\linewidth]{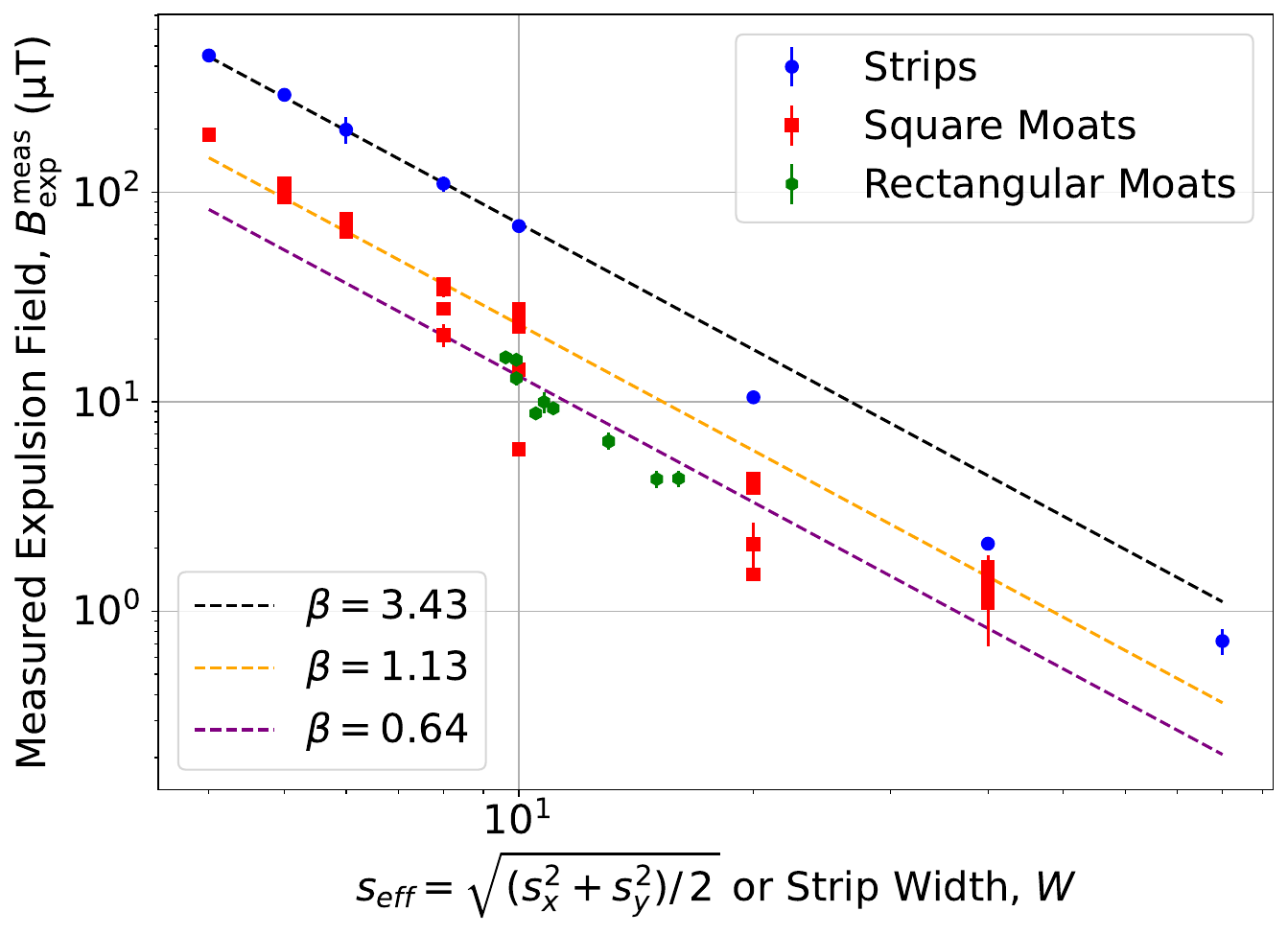}
    \end{overpic}

    \caption{\textcolor{black}{Measured expulsion field vs. effective spacing $s_{\mathrm{eff}}$ for square ($s_{\mathrm{eff}}=s$) and rectangular moat arrays, and vs. strip width $W$ for strips~\cite{qswift_apparatus_paper}. The differences in $B^{\mathrm{meas}}_{\mathrm{exp}}$ for fixed $s_{\mathrm{eff}}$ are related to differences in moat size/shape.}
    }
    \label{fig:basic_s_squared_comparison_asymmetric}
\end{figure}

\textcolor{black}{Fig.~\ref{fig:basic_s_squared_comparison_asymmetric} shows $B^{\mathrm{meas}}_{\mathrm{exp}}$ plotted as a function of the effective moat spacing $s_{\mathrm{eff}}$ for all measured moat arrays. Also shown are the Nb strip data from~\cite{qswift_apparatus_paper}, plotted as a function of strip width $W$. The square moat data are well described by a fit with $\beta_{\mathrm{moat}}=1.13$, except for the smallest 0.5- and 1-$\upmu$m square moats with the largest 40-$\upmu$m spacing. Compared to strips with width $W=s$, square moat arrays exhibit expulsion fields that are smaller by approximately a factor of 3, implying an equivalent strip width of $W\approx\sqrt{3}\,s_{\mathrm{eff}}$.}

\textcolor{black}{More generally, when plotted as a function of $s_{\mathrm{eff}}$ defined in Eq.~\ref{eq:Bexp_slits_2d}, all moat-array data, including long slits, collapse onto a common dependence. This indicates that the expulsion field is governed primarily by the effective moat spacing, $s_{\mathrm{eff}}$, while the dependence on moat size and shape, visible as the vertical spread of data points at a given spacing in Fig.~\ref{fig:basic_s_squared_comparison_asymmetric}, is relatively weak. We next examine these moat-size effects starting from the simpler case of square moat arrays.}

\begin{figure}[htbp]
    \centering 
    \vspace{0.5em} \begin{overpic}[width=0.95\linewidth]{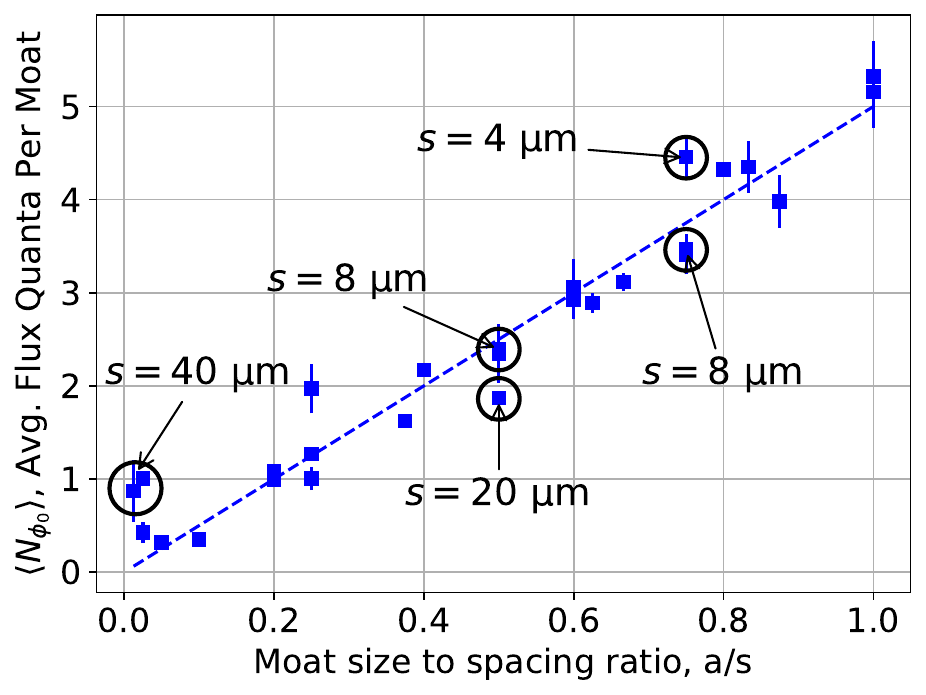} 
    {\footnotesize\textbf{(a)}} \end{overpic} \caption{\textcolor{black}{$\langle N_{\Phi_0}\rangle$, the average amount of flux per moat, vs.~the moat size to spacing ratio for the measured square moat arrays. The data show good agreement with a linear dependence of $\langle N_{\Phi_0}\rangle$ on $a/s$, whereas dependence of $\langle N_{\Phi_0}\rangle$ on ($a/p$) is nonlinear. A linear fit gives $\langle N_{\Phi_0}\rangle= (4.99\pm0.13)(a/s)$. The observed differences in $\langle N_{\Phi_0}\rangle$ at fixed $a/s$ values can be attributed to  differences in $s$ across data points (see, e.g., the difference in $s$ for data points marked by black circles).}} \label{fig:square_moat_size_dependence} 
\end{figure}

$\langle N_{\Phi_0}\rangle$ extracted from the data using Eq.~\ref{eq:Bexp_basic} is plotted in Fig.~\ref{fig:square_moat_size_dependence} versus $a/s\equiv$  \textcolor{black}{($a/s_{\mathrm{eff}}$)}, for all measured square moat, $a_x=a_y=a$, arrays. Although, from flux quantization, individual moats must contain integer number of flux quanta, the array-averaged value $\langle N_{\Phi_0} \rangle$ can deviate slightly due to expulsion via film edges (finite-size array effects) and pinning/nucleating of vortices on film defects. Nevertheless, the data clearly cluster near integer values of $\langle N_{\Phi_0}\rangle$, indicating that these deviations are small.

Figure~\ref{fig:square_moat_size_dependence} shows that arrays with large $a/s$ trap multiple flux quanta per moat, whereas moats in small $a/s$ arrays are only singly or even partially occupied before vortices enter the film. 
$\langle N_{\Phi_0}\rangle$ increases with $a/s$, suggesting the use of closely spaced, large moats for flux sequestration, albeit at the cost of reduced useful circuit area and increased moat--circuit coupling (see Appendix~\ref{subsec:isolated_moats}). \textcolor{black}{A linear fit of the data in Fig.~\ref{fig:square_moat_size_dependence} is shown by the dashed line and gives $\langle N_{\Phi_0}\rangle = (4.99\pm0.13)(a/s)$. Substituting this dependence in Eq.~\ref{eq:Bexp_basic} yields
\begin{equation}
    B_{\mathrm{exp}}^{\mathrm{meas}} \approx 5\Phi_0 (\frac{a}{s})n_{\mathrm{moat}}=5\frac{\Phi_0}{s^2} \frac{a/s}{(1+a/s)^2},
    \label{eq:Bnorm}
\end{equation}} 
\textcolor{black}{which reproduces the governing $s^{-2}$ dependence since $n_{\mathrm{moat}}=p^{-2}=s^{-2}(1+a/s)^{-2}$ for square moat arrays.}

\textcolor{black}{We can extend this approach to rectangular arrays of rectangular moats by using $s_{\mathrm{eff}}$ instead of $s$, and an appropriate parameter $a_{\mathrm{eff}}$, the effective moat side length, instead of $a$. As discussed in Appendix~\ref{subsec:isolated_moats}, the amount of flux that a moat can trap is proportional to the moat perimeter, $P$, which is directly proportional to $a$ for square moats. For the rectangular moats, the equivalent parameter is $a_{\mathrm{eff}}=(a_x+a_y)/2$.} \textcolor{black}{This generalizes Eq.~\ref{eq:Bnorm} to:}

\begin{equation}
    B_{\mathrm{exp}}
    =
    \frac{\gamma\Phi_0}{(a_{\mathrm{eff}}+s_{\mathrm{eff}})^2}
    \frac{a_{\mathrm{eff}}}{s_{\mathrm{eff}}}
    \equiv
    \frac{\gamma\Phi_0}
    {s_{\mathrm{eff}}^2\left(1+a_{\mathrm{eff}}/s_{\mathrm{eff}}\right)^2}
    \frac{a_{\mathrm{eff}}}{s_{\mathrm{eff}}},
    \label{eq:Bnorm_generalized_a}
\end{equation}
\textcolor{black}{demonstrating the $s_{\mathrm{eff}}^{-2}$ behavior plotted in Fig.~\ref{fig:basic_s_squared_comparison_asymmetric}.}
\textcolor{black}{Here $\gamma$ is a dimensionless constant, inferred to be $\approx 5$ from Fig.~\ref{fig:square_moat_size_dependence}.}

\begin{figure}[htbp]
    \centering 
    \vspace{0.5em} \begin{overpic}[width=0.95\linewidth]{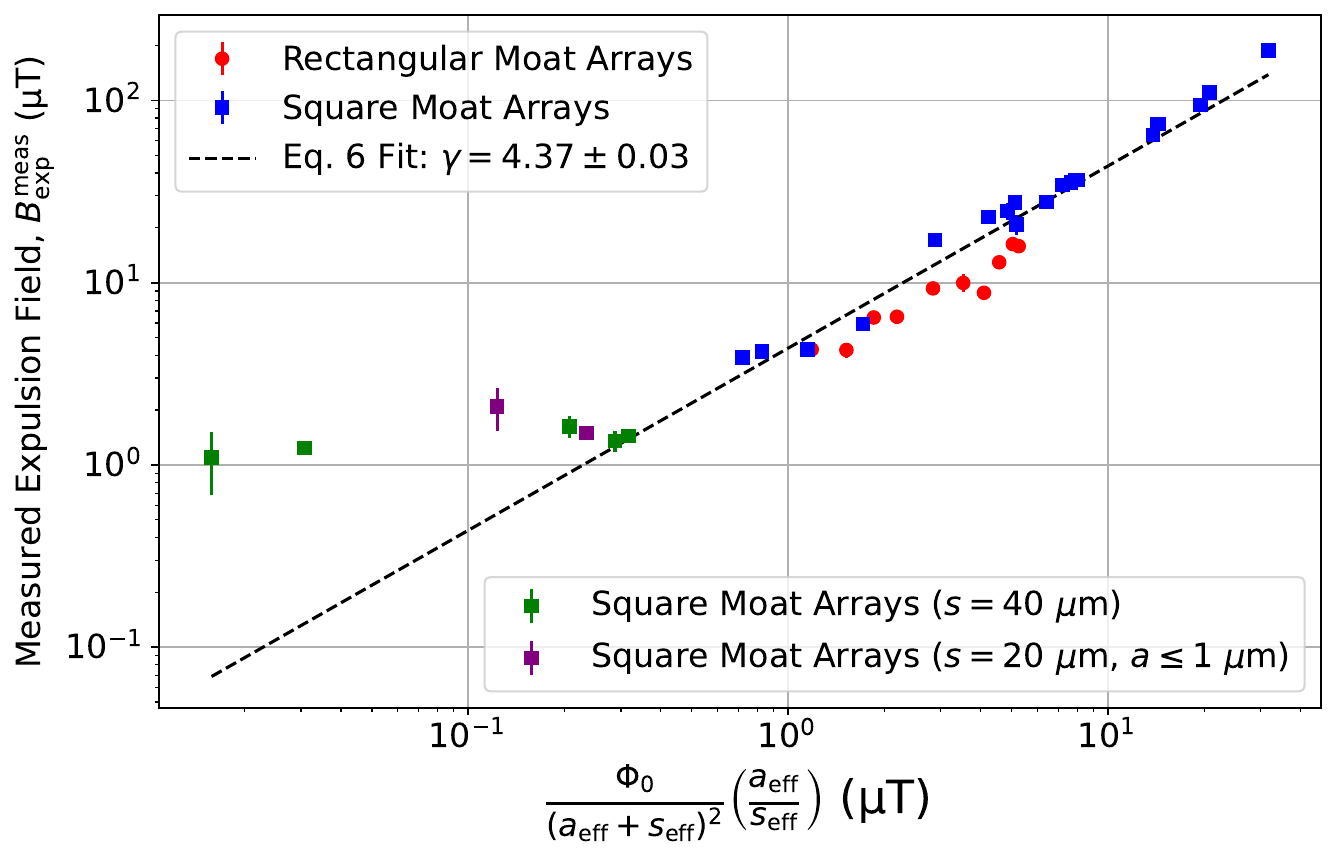} 
    \end{overpic} 
    \caption{\textcolor{black}{$B_{\mathrm{exp}}^{\mathrm{meas}}$ vs. $ 
    \frac{\gamma\Phi_0}{(a_{\mathrm{eff}}+s_{\mathrm{eff}})^{2}}
    \left(\frac{a_{\mathrm{eff}}}{s_{\mathrm{eff}}}\right)$, our proposed expulsion field scaling model in Eq.~\ref{eq:Bnorm_generalized_a}, with $\gamma$ as a fitted parameter, yielding $\gamma=4.37\pm0.03$. The model shows good agreement with the data except for large moat spacing ($s=40$ $\upmu$m and $s=20$ $\upmu$m with $a\leq1$ $\upmu$m, marked in green and purple).}} 
    \label{fig:full_model} 
\end{figure}

\textcolor{black}{In Fig.~\ref{fig:full_model}, all the experimental data on $B_{\mathrm{exp}}^{\mathrm{meas}}$ for square and rectangular moat arrays, including narrow slits, are compared with  Eq.~\ref{eq:Bnorm_generalized_a}, showing very good agreement at $\gamma=4.37\pm0.03$, a 13\% lower value than the $\gamma\approx5$ inferred from Fig.~\ref{fig:square_moat_size_dependence}. The fitted \(\beta_{\mathrm{moat}}\) in Fig.~\ref{fig:basic_s_squared_comparison_asymmetric} is dominated by large moats with \(a\sim s\), which exhibit the largest expulsion fields at a given spacing. As a consistency check, for \(a\lesssim s\) and $\gamma=4.37$, the overall numerical coefficient in front of the $s_{\mathrm{eff}}^{-2}$ dependence in Eq.~\ref{eq:Bnorm_generalized_a} becomes $\approx4.37/4=1.09$, which is very close to the fitted $\beta_{\mathrm{moat}}=1.13$ for square moat arrays in Fig. 5.}

This simple model breaks down at the largest moat spacing, $s = 40~\upmu\mathrm{m}$, where all arrays have $B_{\mathrm{exp}}^{\mathrm{meas}} \lesssim 2~\upmu\mathrm{T}$, comparable to the $s = 20~\upmu\mathrm{m}$ arrays with $a \leq 1~\upmu\mathrm{m}$. At $s = 40~\upmu\mathrm{m}$, $B_{\mathrm{exp}}^{\mathrm{meas}}$ also shows little dependence on moat size $a$ ($1.24\pm0.07~\upmu\mathrm{T}$ and $1.44\pm0.09~\upmu\mathrm{T}$ for $a=1$ and $30~\upmu\mathrm{m}$, respectively), \textcolor{black}{whereas the difference is predicted to be a factor of $\approx10$ from Eq.~\ref{eq:Bnorm_generalized_a}}. This is unexpected, since capturing vortices between widely spaced small moats should become progressively more difficult as the vortex--moat interaction diminishes with distance and decreasing moat size. Instead, we observe a plateau in $B_{\mathrm{exp}}^{\mathrm{meas}}$ versus moat size at $s = 40~\upmu\mathrm{m}$, also visible in Fig.~\ref{fig:square_moat_size_dependence} at very small $a/s$.

\textcolor{black}{The observed plateau in the expulsion field suggests that the smallest \(a\leq1\,\upmu\mathrm{m}\) moats can still trap some amount of flux even at very large spacings \(s\geq20~\upmu\mathrm{m}\), as evidenced by a partial \(\langle N_{\Phi_0}\rangle<1\) filling of the moats at small \(a/s\) for $s=40$ $\upmu$m in Fig.~\ref{fig:square_moat_size_dependence}. However, the expulsion field exhibits a much weaker dependence on moat spacing than for larger moats. The onset of this behavior is likely related to small $a \leq1\, \upmu$m moats losing ability to trap flux at large spacing near \(T_c\) (see Appendices~\ref{subsec:isolated_moats} and \ref{subsec:moat_recs}) and finite size of the moat arrays. The \(s=40~\upmu\mathrm{m}\) arrays studied are the largest by Nb film area, but contain the smallest number of moats due to the largest spacing between them, consisting of only a \(4\times4\) moat array. For comparison, the \(s=20~\upmu\mathrm{m}\) arrays contain \(9\times9\) moats, whereas all arrays with \(s<20~\upmu\mathrm{m}\) contain \(14\times14\) moats. As a result, for the \(s=40~\upmu\mathrm{m}\) arrays, a substantial fraction of the film can expel flux toward the film edges rather than into moats (up to approximately 30--40\%, depending on the moat size). In this regime, flux expulsion to the film edges artificially increases the measured expulsion field and, accordingly, the calculated moat flux filling factor \(\langle N_{\Phi_0}\rangle\).}

Apart from the $s=40$ $\upmu$m moats, both the measured square and rectangular moat arrays show very similar scaling behavior in agreement with our proposed model with only a slight difference in the global offset, likely due to a larger overall size of the rectangular moat arrays. Overall, the data and model indicate that closely spaced moats with large perimeters and aspect ratios provide the highest expulsion field values. 

\textcolor{black}{The superiority of large-perimeter moats, e.g., long slits, is related to the extended film boundaries they introduce, which strongly perturb vortex currents. Qualitatively, the attraction of a vortex to a moat, or more generally to any non-superconducting incursion or film boundary, arises from a disruption of the circular symmetry of the superconducting currents forming the vortex. Because the supercurrent component perpendicular to the moat boundary must vanish, the resulting current redistribution leads to a higher supercurrent density and superfluid velocity in the region between the vortex and the moat compared to the opposite side of the vortex. This enhanced velocity results in a lower Bernoulli pressure in the superfluid between the vortex and the moat, creating a pressure gradient that drives the vortex toward the moat (or film boundary). The magnitude of this attractive force is proportional to the degree of symmetry disruption in the supercurrent flow.}

\textcolor{black}{Small moats with sizes $a \ll \Lambda(T_{\mathrm{exp}})$ produce only a weak perturbation and therefore a small attraction force, since the circulating currents can largely flow around the moat; here $\Lambda(T_\mathrm{exp})$ is the Pearl screening length $2\lambda/d$ at the flux expulsion temperature; see Appendix D. In contrast, long slits with length $l \gg \Lambda(T_{\mathrm{exp}})$ produce a much stronger attraction by confining the supercurrent predominantly between the vortex and the moat. This explains the superiority of long slit-type moats over square moats. An equivalent description can be given in terms of image vortices of opposite sign placed behind the superconducting boundaries~\cite{buzdin1993multiple, buzdin1994theory,Schmidt1974Vortices,nordborg2000interaction,koganringsBc}}.

\textcolor{black}{A moat with trapped flux has a screening current circulating it in the same direction as vortex currents of the same flux polarity. For a moat-vortex pair, the moat current is opposite to the vortex current on the moat side facing the vortex and in the same direction on the far side of the moat. As a result, the average superfluid velocity in the region between the vortex and the moat decreases, decreasing the attraction force with respect to the empty moat or even making the moat repulsive. This explains why we observed $N_{\Phi_0}=1$ for a wide range of square moat sizes. However, if distance between the vortex and the moat, $r$, is much smaller than $\Lambda(T)$, most of the vortex current has to flow around the moat, creating a region of higher superfluid velocity (lower Bernoulli pressure) on the far side of the moat that still attracts the vortex. That is, the vortex-moat interaction may change sign depending on the relations between $a$, $\Lambda$, and $r$ and the moat flux filling factor. The moat current decreases with increasing the moat size, being inversely proportional to the moat inductance, eventually allowing large moats to attract more than one flux quantum; see Appendix~\ref{subsec:isolated_moats}.}
\section{Conclusion}
Magnetic flux trapping remains a critical hurdle in preventing superconducting electronics from reaching very large scale integration. Moats represent one of the simplest practical flux mitigation strategies, requiring only etched patterns in superconducting films and no active circuitry. We have evaluated the effectiveness of various moat geometries in removing magnetic flux by measuring the vortex expulsion field of Nb films with patterned arrays of moats in a wide range of moat sizes, spacings, and aspect ratios. We have found that arrays of closely-spaced, large-perimeter moats, such as long slits or large squares, provide the highest expulsion field values, consistent with the scaling described by Eq.~\ref{eq:Bnorm_generalized_a}. These geometries can trap multiple flux quanta per moat before vortices enter the film and can attract vortices over distances comparable to the moat dimensions. In contrast, sparse arrays of small square (or rectangular) moats ($a_x,a_y\lesssim 5\,\upmu\mathrm{m}$ with spacings $s \ge 2a$) typically trap only a single flux quantum per moat before vortices appear in the film, resulting in substantially lower expulsion fields.

For superconducting integrated circuits, the data suggest the use of narrow slit-type moats with densities commensurate with the expected background field and circuit layout constraints. For a given moat density, slit-type moats can trap more flux quanta than square moats, while also occupying less circuit area, enabling high expulsion fields without excessive loss of usable circuit area. For circuit designs in the SFQ5ee process using 200-nm-thick Nb films, we recommend a moat density of $n_\mathrm{moat}\geq B_r / \Phi_0$, and the maximum moat spacing of less than approximately $16$~$\upmu$m; see Sec.~\ref{subsec:moat_recs} for details. For any designed array and moat geometry, the vortex expulsion field can be estimated using Eq.~\ref{eq:Bnorm_generalized_a} with $\gamma\approx4.4$.  Among the measured geometries, the $a_x = 9 \, \upmu\mathrm{m}, a_y = 0.3\, \upmu\mathrm{m}, s_x = 1\,\upmu\mathrm{m}, s_y = 14\,\upmu\mathrm{m}$ array (see Fig.~\ref{fig:exp_field_measurements}(c)) provides an attractive balance between expulsion field ($>10~\upmu$T) and footprint (1.8\% of a unit cell). The $a_x = 36 \, \upmu\mathrm{m}, a_y = 1\, \upmu\mathrm{m}, s_x = 4\,\upmu\mathrm{m}, s_y = 13\,\upmu\mathrm{m}$ geometry (also used in~\cite {shregNew, shregOld, ac_power_sfq}) offers even higher expulsion field (16.3 $\upmu$T vs.~13 $\upmu$T) at the cost of increased area (6.4$\%$ of the unit cell). Further exploration of high-aspect-ratio slit geometries is therefore well motivated, particularly in regimes approaching their practical limits, while remaining commensurate with circuit design constraints.

For circuits operating in magnetically shielded environments ($B_r\lesssim$ 1 $\upmu$T), most moat geometries are capable of sequestering the majority of flux expelled during cooldown. However, the data show that \textit{moats alone are insufficient to eliminate all trapped flux in non-ideal films}. Whereas the measured expulsion field $B_{\mathrm{exp}}^{\mathrm{meas}}$ characterizes the geometry-dependent threshold at which vortex density $n_v(B_r)$ begins to grow proportionally to  $B_r$/$\Phi_{0}$, across all geometries we have observed a few isolated vortices appearing at lower fields, always at reproducible locations across multiple cooldowns, consistent with nucleation and/or pinning at film defects. An example of this can be seen in Fig.~\ref{fig:exp_field_measurements}(c) for the array with $a_x = 9\,\upmu\mathrm{m}, a_y = 0.3\,\upmu\mathrm{m}, s_x = 1\,\upmu\mathrm{m}, s_y = 14\,\upmu\mathrm{m}$, which exhibits $B_{\mathrm{exp}}^{\mathrm{meas}} > 10$ $\upmu$T, but a few vortices are still observed in the film at $B_r < 1$ $\upmu$T.  Thus, while moats are essential in helping circuits to expel most of the initially threading flux, material defects in non-ideal films may often prevent the complete flux elimination in moat-guarded circuits.


Therefore, mitigating parasitic flux trapping requires optimization of both the moat design and materials used in superconducting electronics, especially in wide ground planes. The emphasis should be on increasing the field at which vortices become trapped on pinning sites and can no longer be sequestered into the moats (field $B_1$ in this work), i.e., on reducing strong vortex pinning but without compromising the critical currents required for circuit operation. Future work will investigate the material dependence of $B_{\mathrm{exp}}^{\mathrm{meas}}$ and $B_1$, the role of film defects and other sources of flux trapping, and additional flux mitigation strategies. A comprehensive solution will likely require the combined optimization of magnetic shielding, moat geometry, superconducting film materials and fabrication, and potentially active circuitry to remove trapped flux.

\section{Acknowledgements}
We thank Tom Osadchy 
for assistance with diamond growth; Peter O'Brien, Jon Wilson, and Matthew Ricci for help with diamond coating; and Anil Mankame, Tom Grasso, George Haldeman, Nate O'Connor, John Cummings, Neel Parmar, Ryan Johnson, and Andrew Maccabe for assistance with instrument design, assembly, and testing. We are grateful to Ravi Rastogi and David Kim for overseeing fabrication of the Nb films and test structures used in this work, and to Rabindra Das for help with packaging. We thank Andrew Wagner, Neel Parmar, Michele Kelley, Joe Belarge, and Logan Bishop-Van Horn for helpful discussions. Rohan Kapur thanks John Kim for continued assistance. 

This material is based upon work supported under Air Force Contract No. FA8702-15-D-0001 or FA8702-25-D-B002. Any opinions, findings, conclusions or recommendations expressed in this material are those of the authors and do not necessarily reflect the views of the U.S. Air Force. Notwithstanding any copyright notice, U.S. Government rights in this article are defined by DFARS 252.227-7013 or DFARS 252.227-7014 as detailed above. Use of this article other than as specifically authorized by the U.S. Government may violate any copyrights that exist in this article. The U.S. Government is authorized to reproduce and distribute reprints for Governmental purposes notwithstanding any copyright annotation thereon.

\appendix       
\setcounter{section}{0}

\section{Areal Vortex Density and Definition of Expulsion Field}
\label{subsec:exp_field_slope}
\textcolor{black}{The purpose of the expulsion field measurements presented in this paper was to measure the magnetic field at which moat arrays are no longer effective at trapping all flux induced by external field $B_r$ during field cooling, indicated by growing areal number density of vortices in the film between the moats. As discussed in the main text, the first vortices in the film are detected at some field $B_1$. They appear to either nucleate or pin on film defects, evidenced by consistent appearance at these sites across multiple temperature cycles. The number of these vortices slowly increases with $B_r$ giving rise to a convex shape of $n_v(B_r)$ dependence. We do not use $B_1$ as the expulsion field for a given moat array because it is sensitive to defects in the film. Instead, we define the array expulsion field $B_{\mathrm{exp}}^{\mathrm{meas}}$ as the magnetic field corresponding to the onset of a much faster (quasi)-linear increase in the vortex areal density with $B_r$, characterizing the field at which a significant amount of flux starts entering the film. This field likely better characterizes ideal, defect-free films for which we expect $B_{\mathrm{exp}}^{\mathrm{meas}}= B_{1}^{\mathrm{ideal}}$. Note, $B_1$ appears to scale roughly with $B_{\mathrm{exp}}^{\mathrm{meas}}$ with arrays with large $B_1$ values exhibiting larger, but similar $B_{\mathrm{exp}}^{\mathrm{meas}}$ values.}

\textcolor{black}{In large-area, type-II thin superconducting films in perpendicular magnetic field, $B_\mathrm{exp}=0$ and the vortex density increases linearly with $B_r$ with a slope of $m\Phi_0^{-1}$, where $m=1$. That is, all applied flux above the expulsion field forms as vortices in the film (see, e.g., \cite{qswift_apparatus_paper}).}

\textcolor{black}{Behavior similar to that observed in the moat arrays is also present in Nb thin-film strips. The expulsion field in Nb strips of width $W$ is non-zero and has a $W^{-2}$ dependence \cite{initial_sc_strip_paper, qswift_apparatus_paper}. The vortex density also initially increases very slightly, likely due to film defects/pinning sites, before increasing sharply with a slope $m \Phi_0^{-1}$. In Nb strips $m\neq1$ has also been observed, and typically $0.7 <m < 1$~\cite{initial_sc_strip_paper, qswift_apparatus_paper}. A $m<1$ slope in Nb strips implies that, unlike bare films, not all applied flux above the expulsion field forms vortices in the film but some fraction of it continues to be expelled. Note that YBCO strips demonstrate similar $0.9<m<1$ behavior~\cite{kuit2009vortex}, but do not show the pinning regime (i.e. $B_{\mathrm{exp}}^{\mathrm{meas}}\approx B_1$), consistent with epitaxial thin-film YBCO samples not containing strong pinning sites~\cite{ wells2015analysis}.} 

\textcolor{black}{It is likely that $m\rightarrow1$ for very large  $B_r$ fields. However, finding this regime by vortex counting and  extrapolating down to define an expulsion field becomes difficult, as single-vortex resolution becomes harder at large $n_v$. In addition, calculating $B_{\mathrm{exp}}^{\mathrm{meas}}$ by extrapolating the slope from larger $B_r$ measurements may not necessarily capture the desired characteristic of $B_{\mathrm{exp}}^{\mathrm{meas}}$ as the field at which all applied flux is no longer expelled from the film.} 

\textcolor{black}{In general, the expulsion field of strips as a function of $W$ is not very well understood, and agreement between experiment and theory varies significantly. Kuit et al.~\cite{kuit2009vortex} provides a very good discussion. Bronson, Gelfand, and Field~\cite{PhysRevB.73.144501} discuss Nb strip expulsion field and vortex density measurements performed in part by Field in~\cite{initial_sc_strip_paper} and show that there is significant disagreement in the vortex density scaling behavior between experiments and the theory based on London approximation. We defer further discussion to these papers and references therein.}

\begin{figure}[htbp]
    \centering
    \vspace{0.5em}
    \begin{overpic}[width=0.95\linewidth]{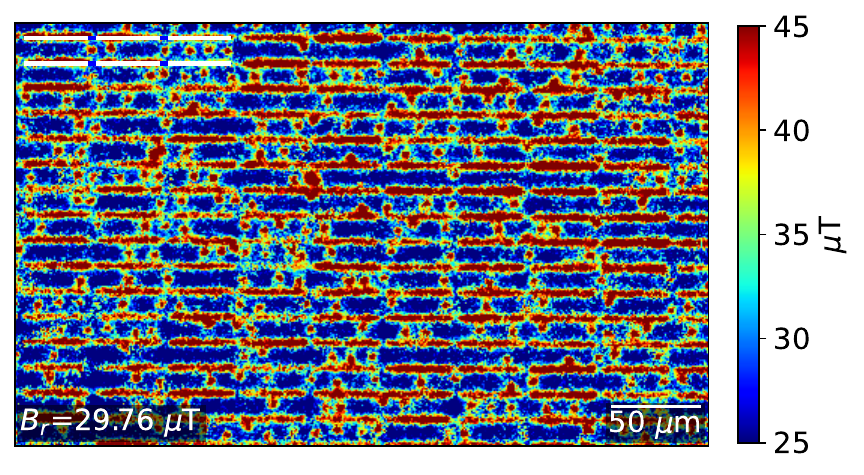}
    \put(-5,48){\footnotesize\textbf{(a)}}
    \end{overpic}
    \vspace{0.5em}
    \begin{overpic}[width=0.95\linewidth]{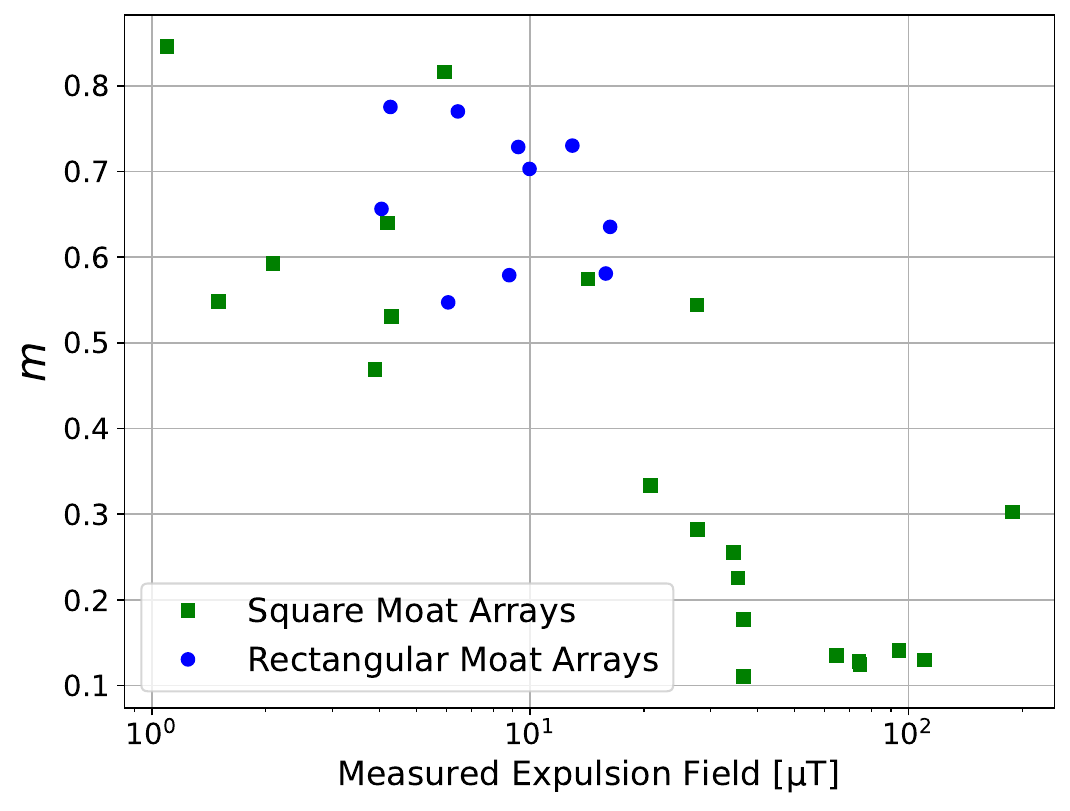}
    \put(-5,68){\footnotesize\textbf{(b)}}
    \end{overpic}

    \caption{\textcolor{black}{(a) Magnetic field image of a moat array with $a_x=36~\upmu$m, $a_y=1~\upmu$m, $s_x=4~\upmu$m, and $s_y=13~\upmu$m for $B_r=$ 29.76 $\upmu$T to be compared with Fig.~\ref{fig:asym_examples}d. While $B_r>B_{\mathrm{exp}}^{\mathrm{meas}}$ for both images, the magnetic field intensity of the moats and amount of flux in the film in Fig.~\ref{fig:high_field_moat5} is significantly higher, indicating that, with $B_r$ increasing above $B_{\mathrm{exp}}^{\mathrm{meas}}$, flux continues to be expelled and trapped into the moats in addition to forming vortices in the film. (b) Slope $m$ of the $n_v=m\Phi_0^{-1}(B_r-B_{\mathrm{exp}}^{\mathrm{meas})})$ dependence as a function of $B_{\mathrm{exp}}^{\mathrm{meas}}$. In bare superconducting films, $B_{\mathrm{exp}}^{\mathrm{meas}}=0$ and $m=1$. For all measured arrays, we observe the slope $m<1$, increasing with decreasing $B_{\mathrm{exp}}^{\mathrm{meas}}$. }} 
    \label{fig:high_field_moat5}
\end{figure}

\textcolor{black}{In the measured moat arrays, we observe behavior very similar to that of the Nb strips. However, $m$, on average, is smaller, being closer to $m\approx0.5$ and varying between $0.1<m<0.8$. This suggests that, above $B_{\mathrm{exp}}^{\mathrm{meas}}$, vortex entry is favored, but flux also continues to be trapped in moats as can also be seen in the magnetic images showing increasing field intensity in the moats. Fig.~\ref{fig:high_field_moat5}a shows the $a_x=36~\upmu$m, $a_y=1~\upmu$m, $s_x=4~\upmu$m, and $s_y=13~\upmu$m moat array, presented before in Fig.~\ref{fig:asym_examples}c-d, at a higher field $B_r=$ 29.76 $\upmu$T. While $B_r>B_{\mathrm{exp}}^{\mathrm{meas}}$ for both images, compared to Fig.~\ref{fig:asym_examples}d, the magnetic field intensity in the moats and amount of flux in the film in Fig.~\ref{fig:high_field_moat5}a is significantly higher, indicating that flux continues to be trapped in the moats at $B_r>B_{\mathrm{exp}}^{\mathrm{meas}}$.}

\textcolor{black}{The value of $m$ depends on the moat array geometry and its flux trapping behavior. In Fig.~\ref{fig:high_field_moat5}b, the slope $m$ is plotted as a function of $B_{\mathrm{exp}}^{\mathrm{meas}}$. As the expulsion field increases, the slope $m$ decreases. This makes intuitive sense; above $B_{\mathrm{exp}}^{\mathrm{meas}}$, the moats continue to trap some flux, and the trapping efficiency should be better in moats with larger $B_{\mathrm{exp}}^{\mathrm{meas}}$ (i.e., a better ability to trap flux). }

\textcolor{black}{As in the strip case, it is possible that $m\rightarrow1$ at $B_r\gg B_{\mathrm{exp}}^{\mathrm{meas}} $. However, we were unable to access such a regime because, at high enough vortex densities, the spatial resolution of our microscope inhibits our ability to count and distinguish between individual vortices. In all measurements presented in the main text, $B_{\mathrm{exp}}^{\mathrm{meas}}$ was extrapolated based on measurements where $n_v\geq 500$ vortices/$\mathrm{ mm}^2$. Although the slope may change beyond the measured density range, the vortex densities used to extrapolate $B_{\mathrm{exp}}^{\mathrm{meas}}$ likely exceed those compatible with reliable circuit operation. For example, shift registers can fail at fields as low as $1\,\upmu\mathrm{T}$, corresponding to a vortex density of $\leq 500$ vortices/mm$^2$~\cite{shregNew}, and likely significantly lower, as these circuits incorporate moats, where $B_{\mathrm{exp}}^{\mathrm{meas}}\gg$ 1 $\upmu$T.}

\textcolor{black}{Finally, in all studied cases, the $n_v (B_r)$ dependence is convex.  This behavior strongly resembles the $n_v(B_r)$ dependence observed in thin-film strips~\cite{initial_sc_strip_paper, qswift_apparatus_paper, maksimova1998mixed}. For thin-film strips without vortex pinning, a simplified model~\cite{maksimova1998mixed} predicts equilibrium vortex areal density $n_v=B_r[1-(B_\mathrm{exp}/B_r )^{1/2}]/\Phi_0$, which is proportional to $B_r/\Phi_0$ at $B_r\gg B_\mathrm{exp}$ and $(B_r-B_\mathrm{exp})/(2\Phi_0)$ near the expulsion field $B_\mathrm{exp}$. A more detailed theory~\cite{PhysRevB.73.144501} accounting for intervortex interactions in $\Lambda >>W $ strips, results in a concave shape of $n_v (B_r)$ dependence describing various ordered configuration of free vortices in the strip and lying between the Maksimova$'$s dependence~\cite{maksimova1998mixed} and $ n_v=B_rW^2/\Phi_0$. In the studied Nb strips and moat arrays, we have not observed ordered lattices of vortices considered in~\cite{PhysRevB.73.144501}, apparently due to the presence of a large number of strong vortex pinning centers in the films with pinning strength significantly exceeding the intervortex interactions responsible for the formation of ordered structures. Including a distribution of pinning centers in the strips into the model in~\cite{PhysRevB.73.144501} with multiple fitting parameters changes the $n_v (B_r)$ dependence towards a more linear dependence observed experimentally in~\cite{initial_sc_strip_paper, qswift_apparatus_paper} but also reduces the value of $B_\mathrm{exp}$ while the experimental data in~\cite{initial_sc_strip_paper, qswift_apparatus_paper}, indicate that $B_\mathrm{exp}$ for narrow strips is a factor of two larger than the theoretical value. Therefore, it is unlikely that the model~\cite{PhysRevB.73.144501} can describe the $n_v (B_r)$ dependence for a more complex geometry of moat arrays. However, we are aware of the preliminary results of the full 3D numerical simulation using time-dependent Ginzburg-Landau (TDGL) equations that correctly reproduce $n_v (B_r)$ dependence observed in this work for various moat geometries~\cite{cadorim_private_comm}.
}
\section{Square Moat Array Measurement Data}
\label{subsec:sq_moat_arrays}
\begin{table}[h]
\centering
\begin{tabular}{|c|c|c|c|}
\hline
$a$ ($\upmu$m) & $s$ ($\upmu$m) & Moat Area & $B_{\mathrm{exp}}^{\mathrm{meas}}$ ($\upmu$T) \\
\hline
3    & 4    & 18.4\% & 188(9) \\
\hline
3    & 5    & 14.1\% & 94(6) \\
\hline
5    & 5    & 25.0\% & 110(8) \\
\hline
4    & 6    & 16.0\% & 64.5(20) \\
\hline
5    & 6    & 20.7\% & 74(5) \\
\hline
6    & 6    & 25.0\% & 74(6) \\
\hline
2    & 8    & 4.0\%  & 21(3) \\
\hline
3    & 8    & 7.4\%  & 27.7(7) \\
\hline
4    & 8    & 11.1\% & 34.4(3) \\
\hline
5    & 8    & 14.8\% & 35.4(13) \\
\hline
6    & 8    & 18.4\% & 36.6(18) \\
\hline
7    & 8    & 21.7\% & 37(3) \\
\hline
1    & 10   & 0.8\%  & 5.93(22) \\
\hline
2    & 10   & 2.8\%  & 17.1(4) \\
\hline
4    & 10   & 8.2\%  & 22.9(6) \\
\hline
6    & 10   & 14.1\% & 24.7(25) \\
\hline
8    & 10   & 19.8\% & 27.62(22) \\
\hline
0.5  & 20   & 0.06\%  & 2.1(6) \\
\hline
1    & 20   & 0.23\%  & 1.50(7) \\
\hline
4    & 20   & 2.8\%  & 3.89(23) \\
\hline
5    & 20   & 4.0\%  & 4.20(19) \\
\hline
10   & 20   & 11.1\% & 4.30(17) \\
\hline
0.5  & 40   & 0.015\%  & 1.1(4) \\
\hline
1    & 40   & 0.06\%  & 1.24(7) \\
\hline
10   & 40   & 4.0\%  & 1.63(22) \\
\hline
20   & 40   & 11.1\% & 1.35(18) \\
\hline
30   & 40   & 18.4\% & 1.44(9) \\
\hline
\end{tabular}
\caption{Geometries and measured expulsion fields of all measured square moat arrays.}
\end{table}

\section{Trapped Flux in an Isolated Moat}
\label{subsec:isolated_moats}
The problem of trapping flux quanta in openings (cylindrical pores, columnar defects, antidots) in superconductors has been studied theoretically and experimentally since the pioneering work by Mkrtchyan and Schmidt \cite{mkrtchyan1972interaction, Schmidt1974Vortices}, who showed that an opening in a superconductor always has an attractive potential for vortices, similar to the interaction with an image antivortex, while the potential of the opening filled with one flux  quanta is attractive at small distances and repulsive at large distances from the opening. They showed that the number of flux quanta at which the potential becomes repulsive at all distances, i.e., the maximum number of flux quanta that can be attracted and trapped into a cylindrical opening with radius $R \ll \lambda(T)$, called the saturation number $N_s$, is given by 
\begin{equation}
N_s(T) = \frac{R^2}{\xi(T)^2 + 2\,\xi(T)\,R},
 \label{eq:saturation_number}
\end{equation} 
which reduces to $R/2\xi(T)$ for $R\gg \xi(T)$, 
where $\lambda(T)$ and  $\xi(T)$ are the temperature-dependent magnetic field penetration depth and coherence length, respectively; see also \cite{buzdin1993multiple, buzdin1994theory, nordborg2000interaction, doria2002multiple} and references therein.  According to Buzdin \cite{buzdin1993multiple}, the $N_s=2$ flux state of a circular hole in the film is energetically stable if 
$R^3 > \xi(T)\,\lambda(T)^2$. In large magnetic fields, the saturation number changes to approximately $R^2/\xi(T)^2$ due to interaction with (pressure created by) adjacent vortices \cite{doria2000maximum}. Trapping of multiple vortices in dense arrays of circular holes, following roughly the latter dependence, was observed, e.g., in \cite{bezryadin1996nucleation}.     Calculations in \cite{buzdin1993multiple, buzdin1994theory, nordborg2000interaction, doria2002multiple,doria2000maximum}, assume that the hole filling by flux quanta occurs sequentially, one by one, and use London or Ginzburg-Landau \cite{ginzburg1950zh} free energy to find the $N_s$ value at which the hole potential changes from attractive to repulsive for the next, $N_{s+1}$, vortex at any vortex-hole separation. This scenario, however, is unlikely to take place during the superconducting transition of a thin film in a weak perpendicular magnetic field because it implies that the Meissner state of the entire film and vortices with currents circulating on a large  $\lambda(T)^2/d$ scale comparable to the spacing between moats and the film width are fully formed at any temperature arbitrarily close to $T_c$. Instead, magnetic flux is probably pushed gradually out from the film simultaneously with the formation of the macroscopic coherent state and circulating currents. 

Regardless of the exact trapping dynamics, the flux saturation number, i.e., the maximum number of flux quanta which can be held in a moat, can be estimated from the fluxoid quantization in superconductors
\begin{equation}
    \upmu_0 \lambda(T)^2 \oint \vec{j}_s \cdot d\vec{l}
    = N_{s}\,\Phi_0 - \oint \vec{A} \cdot d\vec{l},
    \label{eq:fluxoid_quant_theory}
\end{equation}
where $\vec{A}$ is the vector potential of the applied field, $\vec{j}_s$ is the superconducting current density,  and contour can be any closed path in the superconductor. Flux accumulation in the hole (moat) increases the screening current circulating the moat and can proceed until the current density near the moat rim reaches the depairing current density which near the critical temperature is given by the Ginzburg-Landau expression \cite{ginzburg1950zh, ginzburg1958critical,tinkham2004introduction, bardeen1962critical}.
\begin{equation}
    j_c^{\mathrm{GL}} =
    \frac{\Phi_0}{3\sqrt{3}\,\upmu_0\,\xi{(T)}\,\lambda(T)^2}.
    \label{eq:jc}
\end{equation} 
Very close to $T_c$, where the penetration depth is very large, $\lambda(T) \gg d$ and $\lambda(T)^2 /d \gg R,a, s$, and in very small magnetic fields we can neglect the last term $\oint \vec{A} \cdot d\vec{l}$  in Eq.~\ref{eq:fluxoid_quant_theory} in comparison to the first term, i.e., we can neglect magnetic field energy in the moat in comparison to kinetic energy of the supercurrent circulating the moat. Then, for a path around the circular hole rim, the current density $j_s$ is constant and, in this simplified model, $N_s$=$2\pi R/(3\sqrt{3}\xi(T))$ = $1.2R/\xi(T)$, which is just a factor of two larger than the more accurate result of \cite{mkrtchyan1972interaction} at $R\gg \xi(T)$.

For other moat geometries (squares and slits), $j_s$ is not strictly constant around the moat perimeter and can be found by solving London equations in the Meissner state; see \cite{brandt2005thin, mawatari1996critical}. To get an estimate, we neglect corner effects and assume a constant current density around the perimeter of the square moats and long slits. Then, from Eq.~\ref{eq:fluxoid_quant_theory} and Eq.~\ref{eq:jc}, the maximum number of flux quanta that can be held in a moat very close to $T_c$ is
\begin{equation}
    N_s \propto P/\xi(T)
    \label{eq:Ns_vs_perimeter}
\end{equation}  
 where $P$ is the moat perimeter and a proportionality coefficient is $1/3\sqrt{3}=0.19$ for a uniform current distribution. 

Qualitatively the same result follows from a simple electrical engineering formula $\Phi=LI_{\mathrm{cir}}$ where $\Phi$ is the total flux in the moat, $L$ is the moat inductance, and $I_{\mathrm{cir}}$ is the total current circulating the moat. Very near $T_c$, kinetic inductance of the film surrounding the moat $L_k=\upmu_0 \lambda(T)^2 / d$ (per square) is much larger than the moat geometrical inductance, $\simeq \upmu_0a$. Then, the total inductance can be estimated as $L=\upmu_0 (P/w) \lambda(T)^2 /d$ where $P/w$ is the number of squares in the film and $w$ is the typical width of the area in which currents are flowing around the moat; $w \sim s/2$ in arrays. The maximum flux the moat can hold, $\Phi_{max}$, corresponds to the circulating current reaching the critical value $j_c\, d\, w$. Using Eq.~\ref{eq:jc} for $j_c$, we get the same expression
$N_s=\Phi_{max} /\Phi_0=kP/\xi(T)$, where $k$ is a numerical coefficient depending on the moat shape. For a square hole with size $a$, $N_{s}\sim \frac{0.8a }{\xi{(T)}} $, and for a long slit with length $l$, $N_{s}\sim \frac{0.4l }{\xi{(T)}}$. 

Unfortunately, we do not know at what temperature flux expulsion into the moat occurs and, hence, the value of $\xi{(T)}$. Nevertheless, we can assume that flux expulsion into the moats occurs at the same temperature regardless of the shape and density of moats. We also assume that the realized  $N_{\Phi_0}$ value does not change as the film temperature is lowered to the base temperature of the measurements, about 5 K, despite the decrease of $\xi(T)$, because the remaining, not trapped close to $T_c$, flux becomes immobile as frozen vortices and because the moat potential becomes repulsive at large distances from the moat. That is, we assume that the flux distribution imaged in the measurements is the frozen snapshot of the process that happened very close to $T_c$. Then, for sparse arrays, the ratio of $N_{\Phi_0}$ for different moat shapes should be equal to the ratio of their perimeters, which can be easily verified experimentally, e.g., using the data in Table~\ref{tab:rect_moats}. \textcolor{black}{For instance, for the measured arrays of slits with (i) $a_x=36$, $a_y=1$, $s_x=4$ and $s_y=13$ $\upmu$m, (ii) $a_x=19.5$, $a_y=0.3$, $s_x=1$ and $s_y=14$ $\upmu$m, and (iii) $a_x=9$, $a_y=0.3$, $s_x=1$ and $s_y=14$ $\upmu$m, the average $\langle N_{\Phi_0} \rangle= 5,\  2.9$ and $1.15$, respectively, having the ratios $4.3:2.5:1$, which are close to the ratios of their perimeters $4:2.1:1$}. 

However, for all square moat arrays with $a=1$, 2, 3, 4, 5 $\upmu$m and $p_x \geq 20$ or $p_y \geq 20$ $\upmu$m, $N_{\Phi_0}=1$ was found, despite large differences in the moat perimeters; see Fig.~\ref{fig:exp_field_measurements} and~\ref{fig:Bexp_vs_basic_scaling}. That is the measured, actually realized upon field cooling,  $N_{\Phi_0}$ value is much smaller than the flux saturation value $N_s$ characterizing the moat potential capacity. This indicates that square and small-aspect-ratio rectangular moats filled with just one flux quantum become repulsive and $cannot$  $attract$ vortices located at distances comparable to or larger that the moat size $a$, although they potentially could hold more than one flux quantum. However, closely spaced square moats with spacings $s\leq a$ can attract and hold up to 5$\Phi_0$; see Fig.~\ref{fig:Bexp_vs_basic_scaling} and Fig.~\ref{fig:square_moat_size_dependence}. This indicates once again the importance of electromagnetic interactions between vortices and moats, and between the adjacent moats, and confirms their short range nature \cite{buzdin1994theory}, as well as importance of dynamic effects in expelling flux into the moats. For instance, vortices can hardly be attracted from symmetrical positions in the middle between two or four moats, i.e., at distance $a/2$ from the moat side, where the attraction/repulsion forces are balanced. 

\section{Estimate of Flux Trapping Temperature and Circuit Geometry Recommendations}
\label{subsec:moat_recs}
Linear scaling of $N_{\Phi_0}$ with $a/s$ in Fig.~\ref{fig:square_moat_size_dependence} indicates that the moat potential, which at small distances decays approximately as $1/r$ for a single circular or square moat (where $r$ is the distance from the moat center) is significantly modified by the presence of adjacent moats, allowing moats in dense arrays to trap more flux quanta per moat than an isolated moat. 
Empirically, the effective area from which magnetic flux can be swept into the moats of given sizes $a_x$, $a_y$ and spacings $s_x$, $s_y$ is given by the scaling functions in Eq.~\ref{eq:Bnorm_generalized_a}.

Using the data in Fig.~\ref{fig:square_moat_size_dependence}, we can estimate the value of $\xi(T)$  at a temperature when flux expulsion from the film and trapping into the moats occurs, $T_{\mathrm{exp}}$. Indeed, sparse ($s\gg a$) arrays trapped $N_{\Phi_0}=1$ down to sizes $a_{min}$ of about $1 \,\upmu$m. \textcolor{black}{ Square moats with $a=0.5$ $\upmu$m were typically occupied at about 50\%  before vortices appeared between the moats. From Eq.~\ref{eq:saturation_number}, $N_s$=1 at $\xi(T)=R(1+\sqrt{2})\approx2.4R$. If we approximate the square moats as circles with effective radius $R_\mathrm{eff}=2a/\pi$, then $\xi(T_\mathrm{exp}) \approx 4.8a_{min}/\pi \approx 0.76 \upmu$m for $a_{min}$ = 0.5 $\upmu$m and $\approx1.5$ $\upmu$m for $a_{min}$ = 1 $\upmu$m. From Eq.~\ref{eq:Ns_vs_perimeter}, $N_s=1$ at $\xi(T)\approx0.8a_{min}$ for square moats, giving $\xi(T_\mathrm{exp})$ of about 0.4-0.8 $\upmu$m.}

\textcolor{black}{The coherence length in our Nb films can be estimated using the coherence length in clean Nb at $T = 0$, $\xi_0 = 39~\mathrm{nm}$, and the electron mean free path $l = 12~\mathrm{nm}$, obtained from the films' residual resistivity $\rho_0 = 3.2~\upmu\Omega\cdot\mathrm{cm}$ and $\rho l = 3.75 \times 10^{-12}~\Omega\cdot\mathrm{cm}^2$. In the dirty limit $l \ll \xi_0$, the coherence length near $T_c$ is given by $\xi_d(T) = 0.85(\xi_0 l)^{1/2}(1 - T/T_c)^{-1/2}$, or $\approx 18(1 - T/T_c)^{-1/2}$ in nm.} This gives
$T_{\mathrm{exp}}/T_c=1-(18/\xi(T_\mathrm{exp}))^2$ in the range from 0.9980 to 0.9994 depending on the assumed value of $\xi(T)$ (from 0.4 to 0.76 $\upmu$m) as discussed above. So, flux expulsion and trapping in the moats occurs down to $\gtrsim$ 20 mK below $T_c$. At this temperature, the Pearl screening length $\Lambda(T)=2\lambda(T)^2 /d$~\cite{pearl1964current} that sets the range of vortex interactions in thin films is $\Lambda(T_\mathrm{exp})\approx8.1\,\upmu$m in 200-nm thick Nb films with $\lambda(0)=81$ nm~\cite{tolpygo2018superconductor}. 

Using the estimated $\xi(T_\mathrm{exp})$ and $\Lambda(T_\mathrm{exp})$, we can estimate the radius of circular moats that can support $N_s=2$ flux saturation \cite{buzdin1993multiple}  as $R>3.7\, \upmu$m, corresponding to square moat sizes $a>5.8\,\upmu$m, which is in full agreement with our observations that sparse moats with sizes up to $5\, \upmu$m trap just a single flux quantum.     
 
At distances larger than $\Lambda(T)$ currents circulating the vortex decay as $r^{-2}$~\cite{pearl1964current}. Therefore, moats spaced at $s \gg \Lambda(T_\mathrm{exp})$ become ineffective in trapping vortices located between them because similarly quickly decaying with distance attraction force to the moats becomes insufficient to overcome even weak pinning forces created by defects in the film.   In practice, the recommended spacing between the moats should be $s\lessapprox 2\Lambda(T_\mathrm{exp}) \approx16 \,\upmu$m in 200-nm-thick Nb films. It should be scaled proportionally to $d^{-1}$ for thicker or thinner films.

\textcolor{black}{The estimated lower temperature of flux expulsion temperature range $T_{\mathrm{exp}}$ can be also viewed as a vortex freezing temperature, $T_f$ in notations of Stan~\cite{initial_sc_strip_paper}, below which vortices become practically immobile in the random potential of pinning centers in the film and cannot be expelled into the moats despite that the moats may still have an attractive potential and trapping capacity, i.e., the average flux in them $\langle N_{\Phi_0} \rangle$ is still less that the flux saturation number $N_s$.}

\textcolor{black}{The estimates provided above are very sensitive to the selection of $a_{min}$ because both $\xi(T)$ and $\Lambda(T)$ diverge at $T\rightarrow T_c$. For instance, if we use the upper estimate $\xi(T_\mathrm{exp}) \approx$ 0.76 $\upmu$m based on $a_{min}=0.5\,\upmu$m and Eq.~\ref{eq:saturation_number}, then the flux expulsion temperature range would be $T_c-T_{\mathrm{exp}}\approx 5$ mK and $\Lambda(T_\mathrm{exp})\approx$ 28.8 $\upmu$m.  However, this extremely small temperature range lies completely withing the natural width of the superconducting transition of Nb films caused by their nonuniformities, implying questionable applicability of the described models assuming a uniform superconducting state.}

\textcolor{black}{Finally, the estimated Pearl length of about 8 $\upmu$m at the end of the expulsion temperature range is consistent with the expulsion field measurements on long thin-film strips in~\cite{qswift_apparatus_paper}, which show an apparent crossover from $W\ll\Lambda(T)$ to $W>\Lambda(T)$ at the strip widths $W$ about 10 $\upmu$m visible in Fig.~\ref{fig:basic_s_squared_comparison_asymmetric}. }

\bibliography{biblio.bib}

\end{document}